%% file: main.tex
\documentclass[10pt,letterpaper]{article}
\input{header}

\let\oldequation\equation
\let\oldendequation\endequation
\renewenvironment{equation}
  {\linenomathNonumbers\oldequation}
  {\oldendequation\endlinenomath}
\nolinenumbers

\title{Stimulus-to-stimulus learning in RNNs with cortical inductive biases}

\author[1*]{Pantelis Vafidis}
\author[2]{Antonio Rangel}

\affil[1]{Computation \& Neural Systems, California Institute of Technology, Pasadena, CA 91125}
\affil[2]{Humanities and Social Sciences, California Institute of Technology, Pasadena, CA 91125}
\affil[*]{pvafeidi@caltech.edu}

\usepackage[numbers,sort&compress]{natbib}
\usepackage[hidelinks]{hyperref}
\usepackage{verbatim}
\usepackage{cleveref}   % to include words Figure, Equation etc. when referencing

\begin{document}

\maketitle
\renewcommand{\abstractname}{Summary}
\begin{abstract}
Animals learn to predict external contingencies from experience through a process of conditioning. A natural mechanism for conditioning is stimulus substitution, whereby the neuronal response to a stimulus with no prior behavioral significance  becomes increasingly identical to that generated by a behaviorally significant stimulus it reliably predicts. We propose a recurrent neural network model of stimulus substitution which leverages two forms of inductive bias pervasive in the cortex: representational inductive bias in the form of mixed stimulus representations, and architectural inductive bias in the form of two-compartment pyramidal neurons that have been shown to serve as a fundamental unit of cortical associative learning. The properties of these neurons allow for a biologically plausible learning rule that implements stimulus substitution, utilizing only information available locally at the synapses. We show that the model generates a wide array of conditioning phenomena, and can learn large numbers of associations with an amount of training commensurate with animal experiments, without relying on parameter fine-tuning for each individual experimental task. In contrast, we show that commonly used Hebbian rules fail to learn generic stimulus-stimulus associations with mixed selectivity, and require task-specific parameter fine-tuning. Our framework highlights the importance of multi-compartment neuronal processing in the cortex, and showcases how it might confer cortical animals the evolutionary edge.
\end{abstract}

% Our framework complements more common auto-associative memory architectures, by allowing concepts to be associated together.


%\section*{Author summary}
%TBD.

\nolinenumbers
\onehalfspacing

\clearpage

\input{intro}

\input{results} 

\input{discussion}

\input{methods} %notmain

\paragraph{Acknowledgements}
A.R. gratefully acknowledges support from the NOMIS foundation and from NIH grant R01MH134845. P.V. gratefully acknowledges support from the Onassis Foundation. P.V. would like to thank John O'Doherty whose class was an initial inspiration for this project. No competing interests to disclose.

\paragraph{Author contributions}
P.V. conceived the project, developed the model, performed analyses and wrote the initial draft of the manuscript. A.R. supervised the research and provided funding. A.R. and P.V. wrote the final version of the manuscript.

\paragraph{Data and code availability}
All code along with data to replicate the results will be shared upon acceptance.

\nolinenumbers
\bibliographystyle{plos2015}
\bibliography{references}

\input{supplementary} %notmain

\end{document}

%% file: header.tex
\usepackage{graphicx}
\usepackage{csquotes}
\usepackage{amssymb,amsmath}
\usepackage{nicefrac}
\usepackage{subfigure}
\usepackage{authblk}
\usepackage[left]{lineno}
\usepackage{setspace}
\usepackage{placeins}
\usepackage{bbm}

\usepackage{xcolor}  % for \todo
\MakeOuterQuote{"}

\usepackage{booktabs}

\usepackage{geometry}
\usepackage{changepage}
\newlength{\offsetpage}
{\end{adjustwidth}}

%% file: intro.tex
%!TEX root = main.tex

\section*{Introduction}

The ability to forecast important events is necessary for effective behavior. Animals are equipped with innate reflexes to tackle common threats and to exploit opportunities in their environment. However, given the complex and changing nature of the world, animals also need to acquire new reflexes by learning from experience. This process involves the association or conditioning of an initially neutral stimulus (conditioned stimulus, \textit{CS}) with another stimulus intrinsically related to primary reward or punishment (unconditioned stimulus, \textit{US}). If learning is successful, the \textit{CS} can then induce the same behavioral response as the \textit{US}. Initially proposed by Pavlov, this type of learning is known as classical conditioning.

A potential mechanism for conditioning is stimulus substitution \cite{Jenkins1973}. Under this mechanism, the response of the relevant population of neurons to the \textit{CS} becomes increasingly identical to that generated by the \textit{US}. After this, any downstream processes that are normally triggered by the \textit{US} are also triggered by the \textit{CS}. Behavioral evidence in favor of stimulus substitution comes from studies showing that animals display the same behavior to the \textit{CS} as to the \textit{US}, even when the behavior is not appropriate (e.g. consummatory response towards a light that has been associated with food), and that the behavior is reinforcer dependent \cite{Jenkins1973}. Furthermore, recent experiments show that during conditioning the response of S1 pyramidal neurons to the \textit{CS} becomes increasingly similar to their response to the \textit{US}, a phenomenon the authors termed "learning induced neuronal identity switch", and that this change correlates with learning performance \cite{Dai2023}. 

A basic goal in computational and cognitive neuroscience is to build plausible models of neural network architectures capable of accounting for psychological phenomena. Previous work has shown that three-factor Hebbian synaptic plasticity rules accounts for a wide gamut of conditioning phenomena \cite{Sutton1987,Klopf1988,Balkenius1998,Izhikevich2007}. However, these models have some important limitations. First, they fail to capture the generality of pattern-to-pattern associations implicit in stimulus substitution, where both the \textit{US} and the \textit{CS} correspond to population activity patterns. Some use learning rules requiring storage of recent events at each synapse \cite{Balkenius1998}, while most assume that the tuning of neurons to stimuli is demixed, allowing simple reward modulated spike-timing-dependent plasiticy to establish the appropriate mappings \cite{Balkenius1998,Izhikevich2007}. These assumptions are inconsistent with the well-established fact that representations throughout the brain are high-dimensional and mixed \cite{Rigotti2013}. 

In this study we propose a recurrent neural network (\textit{RNN}) model of stimulus substitution. Critically, the model learns pattern-to-pattern associations using only biologically plausible local plasticity, and individual neurons are tuned to multiple behavioral stimuli, which gives rise to mixed representations of the \textit{CS}s and \textit{US}s. While subcortical \cite{Christian2003} and even single-neuron \cite{Gershman2021} mechanisms for conditioning exist, our model is focused on stimulus-stimulus learning in the cortex, where the use of mixed stimulus representation allows learning a wide and flexible range of associations within the same neuronal network, which confers an evolutionary edge. 

To achieve this goal, we leverage two forms of biological inductive bias built into the cortex: first, representational inductive bias in the form of mixed stimulus representations, that permit the efficient packing of multiple associations within the same neuronal population. To combat the additional complexity introduced by mixed representations, which requires not just the activation of the correct neurons but also the correct activity level, we leverage the second form of inductive bias: architectural inductive bias in the form of two-compartment layer-5 pyramidal neurons which are prevalent in the cortex \cite{Nieuwenhuys1994}.

We propose a \textit{RNN} model of such two-compartment neurons. Recent work has shown that these neurons can learn to be predictive of a reward \cite{Doron2020}, and suggests that they could serve as a fundamental unit of associative learning in the cortex through a built-in cellular mechanism \cite{Larkum2013}. Hence, we refer to them as associative neurons. The term associative here does not have a strictly Hebbian interpretation; rather it refers to the \textit{hetero}-associative capacity of these neurons to link together information originating from different streams \cite{Shin2021}, through a mechanism known as BAC firing \cite{Larkum1999}. The properties of these neurons allow for a biologically plausible learning rule that utilizes only information available locally at the synapses, and that is capable of inducing self-supervised predictive plasticity \cite{Urbanczik2014,Urbanczik2009}, which allows neurons to respond with the same firing rate to the \textit{CS} as they would to the \textit{US}, i.e. achieve stimulus substitution.

We show that the model generates a wide array of conditioning phenomena, including delay conditioning, trace conditioning, extinction, blocking, overshadowing, saliency effects, overexpectation, contingency effects and faster reacquisition of previous learnt associations. Furthermore, it can learn large numbers of \textit{CS-US} associations with an amount of training commensurate with animal experiments, without relying on parameter fine-tuning for each individual experimental task. In contrast, we show that Hebbian learning rules, including three-factor extensions of Oja's rule \cite{Oja1982} and the BCM rule \cite{Bienenstock1982}, fail to learn generic stimulus-to-stimulus associations due to their unsupervised nature, and require task specific parameter fine-tuning.

%% file: results.tex
%!TEX root = main.tex

\section*{Results}

\subsection*{Model setup}

In classical conditioning animals learn to predict the upcoming appearance of an unconditioned stimulus (\textit{US}, e.g. food) after the presentation of a conditioned stimulus (\textit{CS}, e.g. bell ring). As shown in \cref{fig:model}A, trials start with the presentation of the \textit{CS}, which lasts until $t_\text{cs-off}$. The \textit{US} is presented at $t_\text{us-on}$, and lasts until the end of the trial. Each trial has a fixed duration of $t_\text{trial}$ seconds. If the \textit{US} appears before the \textit{CS} disappears, the task involves delay conditioning. In contrast, if the \textit{CS} disappears before the \textit{US} is shown, the task involves trace conditioning, with $t_\text{delay} = t_\text{us-on} - t_\text{cs-off}$ denoting the delay between the two stimuli. In our task animals need to learn $N_\text{stim}$ different \textit{CS}-\textit{US} pairs. Every trial one pair is randomly chosen, and the corresponding \textit{CS} is shown followed by its associated \textit{US}.

We model a \textit{RNN} of associative neurons (\cref{fig:model}C, yellow background) that represents the stimuli using mixed population representations and is capable of learning all of the \textit{CS}-\textit{US} associations using only local information available at the synapses. The inputs to the model are time-dependent vectors $r_\text{cs}(t)$ and $r_\text{us}(t)$, of dimension $N_\text{inp}$, that encode the presence and identity of the \textit{CS} and the \textit{US}. For simplicity, these vectors are represented by unique Boolean vectors, and they take the value of the stimulus while it is shown, and zero otherwise. The vectors are randomly generated, subject to a constraint for a minimal Hamming distance $H_\text{d}$ between any two vectors of the same type. This minimal separation limits the extent to which learning on any give pair impairs learning of the other associations. The output of the associative network is an estimate of the \textit{US} vector $r_\text{us}$, denoted $\hat{r}_\text{us}$, which is decoded from network activity at all times (see \cref{fig:model}C and "\textit{US} decoding" in Methods).

%%%%%%%%%%%
\input{figs_text/fig1_text.tex}
%%%%%%%%%%%

The fundamental unit of computation in the associative network is the associative neuron, a two-compartment neuron modelled after layer-5 pyramidal cells in the cortex (\cref{fig:model}B). A crucial property of the associative neuron is that it can separate incoming "feedforward" inputs from "feedback" ones, and compare the two to drive learning. In our case, since we are modelling a primary reinforcer cortical area, \textit{US} inputs are assumed feedforward and arrive at the somatic compartment (corresponding to the soma and proximal dendrites) through synaptic connections  $W_\text{us}$, and \textit{CS} inputs are considered feedback connections arriving to the distal dendrites from the rest of the cortex, along with local recurrent connections ($W_\text{cs}$ and $W_\text{rnn}$ respectively, \cref{fig:model}B). This separation of inputs ultimately allows for the construction of a biologically plausible predictive learning rule, capable of achieving stimulus substitution.

Specifically, to account for the ability of the associative neuron to predict its own spiking activity to somatic inputs from dendritic inputs alone \cite{Larkum1999}, we utilize a synaptic plasticity rule that implements local error correction at the neuronal level \cite{Urbanczik2014}. The learning rule modifies the connections to the dendritic compartment (i.e. $W_\text{cs}$ and $W_\text{rnn}$) in order to minimize the discrepancy between the firing rate of the neuron $f(V^\text{s})$ (where $V^\text{s}$ is the somatic voltage, primarily controlled by \textit{US} inputs in the beginning of learning, and $f$ the activation function) and the prediction of the firing rate by the dendritic compartment $f(p^\prime V^\text{d})$ (where $V^\text{d}$ is the dendritic voltage, primarily controlled by \textit{CS} inputs, and $p^\prime$ is a constant accounting for attenuation of $V^\text{d}$ due to imperfect coupling with the somatic compartment). The synaptic weight $W_\text{pre,post}$ from a presynaptic neuron to a postsynaptic associative neuron is modified according to:
\begin{equation}
\Delta W_\text{pre,post} = \eta (S) \left[\,f(V^\text{s}_\text{post})-f(p^\prime \, V_\text{post}^\text{d})\,\right]\, \textit{P}_\text{pre}
\label{eqn:learn_rule_main}
\end{equation}
where $\eta$ is a variable learning rate which depends on a surprise signal $S$ and $\textit{P}_\text{pre}$ the postsynaptic potential from the presynaptic neuron (for details, see "Synaptic plasticity rule" in Methods). In the Supplementary Information (section "Predictive coding and normative justification for the learning rule") we show how this learning rule can be derived directly from the objective of stimulus substitution.

During trace conditioning the \textit{CS} disappears before the \textit{US} appears, but an association is still learnt. This suggests that the brain maintains some short-term memory representation of the \textit{CS} after it disappears. To capture this, we introduce a short-term memory \textit{RNN} that maintains a (noisy) representation of the \textit{CS}, denoted by  $\hat{r}_\text{cs}$, over time (for details, see "\textit{CS} short-term memory circuit" in Methods). As shown in \cref{fig:model}D, the network is able to maintain short-term representations of the \textit{CS} for several seconds before memory leak becomes considerable.

Finally, the learning rule is gated by a surprise mechanism mediated by diffuse neuromodulator signals \cite{Gerstner2018}, as follows: Upon \textit{CS} presentation, an expectation $E$ is formed according to the proximity of $\hat{r}_\text{us}$ to $r_\text{us}$ for known \textit{US}s (see "\textit{US} expectation estimation" in Methods). $E$ can be thought of as the probability that some known \textit{US} will appear. Upon \textit{US} presentation, $E$ is compared to 1 and a surpise signal $S = 1 - E$ is formed and gates learning; the greater the surprise, the greater the learning rate. If no \textit{US} appears in the trial, then we set $S = - E$ at $t_\text{wait}$ seconds after normal \textit{US} presentation. Non-zero values of $S$ activate learning, in a process driven by two neuromodulators, one for positive and another for negative learning rates (for details on dynamics, see "Surprise based learning rates" in Methods).

% -----------------------------------------------------------
\subsection*{Network learns stimulus substitution in delay conditioning}
% -----------------------------------------------------------

Consider a delay conditioning experiment in which the animal needs to learn 16 \textit{CS-US} pairs, and the timing of the trial is as shown in \cref{fig:delay_cond}A. Note that in this case the \textit{CS} is present throughout the trial and, as a result, $\hat{r}_\text{cs} \approx r_\text{cs}$. Although the short-term memory network is not necessary in this particular experiment, we keep it in the model to maintain consistency across experiments.

We train the \textit{RNN} for a total of 1000 trials. \Cref{fig:delay_cond}B compares the actual representations of all the \textit{US}s, one component at a time, with those decoded from the activity of the network in response only to the associated \textit{CS}s. The network has accurately learnt all of the associations after 500 training trials ($\approx$ 32 per \textit{CS-US} pair).

%%%%%%%%%%%
\input{figs_text/fig2_text.tex}
%%%%%%%%%%%

We next investigate how learning evolves with the amount of training. \Cref{fig:delay_cond}C compares the activity of the associative neurons when presented only with the \textit{US}, for all possible \textit{CS-US} pairs, with their activity when presented only with the associated \textit{CS}. Early in training, the associative neurons exhibit little activity in response to the \textit{CS}s, and their responses are not correlated with the amount of activity elicited by the \textit{US}s. By the end of training however, the neurons respond to the \textit{CS} the same way they respond to the \textit{US}, therefore stimulus substitution is achieved. A host of conditioning phenomena, detailed in following sections, follow from that. For further details on the trial dynamics of learning see the Supplementary Information (section "How does the \textit{RNN} learn?"). Importantly, in the Supplements we also show that three-factor Hebbian learning rules fail at stimulus substitution in our experiments.

\Cref{fig:delay_cond}D tracks the learning dynamics more closely. The green curve shows the average expectation $E$ assigned to the \textit{US}s at different stages of training. Perfect learning occurs when $E=1$ for all \textit{US}s.  The red curve provides a measure of distance between the $r_\text{us}$ and $\hat{r}_\text{us}$. We see that learning requires few repetitions per \textit{CS-US}, and is substantially faster early on.

There are three sources of randomness in the model: (1) randomness in the sampling of \textit{CS} and \textit{US} sets, (2) randomness in the order in which the stimulus pairs are presented, and (3) randomness in the initialization of $W_\text{rnn}$, $W_\text{cs}$ and $W_\text{us}$. In \cref{fig:multiple_runs} we explore the impact of this noise in our results by training 5 networks with different initializations and training schedules. We find that the level of random variation across training runs is small, and is mostly dominated by randomness in the sampling of the stimuli. For this reason, unless otherwise stated, we present results using only a single  training run.

Since the \textit{RNN} uses mixed representations over the same neurons to encode the stimuli, one natural question is how does learning depend on the number of \textit{CS-US} pairs in the experiment ($N_\text{stim}$) and on the similarity of their representations ($r_\text{cs}$ vs $r_\text{us}$). 

We explore the first question by training the model for different values of $N_\text{stim}$ and then measuring the number of trials that it takes the network to reach a 80\% level of maximum performance, defined as the level of training at which the average expectation $E$ across pairs exceeds $0.8$. Interestingly, the required number of trials increases exponentially with the number of \textit{CS-US} pairs (\cref{fig:delay_cond}E). This is likely due to interference across pairs: learning of an association also results in unlearning of other associations at the single trial level. This interference gets worse as the number of stimuli $N_\text{stim}$ increases (\cref{fig:n_stim}), which might explain the exponential dependence. Finally, note that the network is capable of very fast learning when there are only a few pairs (about 5 presentations per pair for two pairs, \cref{fig:delay_cond}E).

We explore the second question by training the model for different values of the Hamming distances $H_\text{d}$, which provides a lower bound on the similarity among \textit{US}s and, separately, among \textit{CS}s. $N_\text{stim}=8$ for these experiments. Perhaps unsurprisingly, the more dissimilar the stimulus representations, the faster the learning (\cref{fig:delay_cond}F). \Cref{fig:hamming} shows how smaller $H_\text{d}$ naturally leads to greater interference across stimuli.

% -----------------------------------------------------------
\subsection*{Short-term memory and trace conditioning}
 % -----------------------------------------------------------

Next we consider trace conditioning experiments, in which there is a delay interval $t_\text{delay}>0$ between the disappearance of the \textit{CS} and the arrival of the \textit{US} (\cref{fig:trace_cond}A). In this case the memory network is crucial for maintaining a memory trace of the \textit{CS} to be associated with the \textit{US}.

As before, we train the \textit{RNN} for 1000 trials, with $16$ different pairs, to explore how learning changes over time and how the delay $t_\text{delay}>0$ affects 
learning. For comparison purposes, we include the case of delay conditioning in the same figures ($t_\text{delay} = -1$ s).

\Cref{fig:trace_cond}B shows the quality of the decoded representation of the \textit{US} and \cref{fig:trace_cond}C-D the strength of the associated expectation signal, both measured offline and in response only to the \textit{CS}. We find that the \textit{RNN} learns the associations well for small delays, but that the quality of the learning decays for larger delays. This pattern has been observed in animal experiments \cite{Schneiderman1964}, and the model provides a mechanistic explanation: conditioning worsens with increasing delays because the memory representation of the \textit{CS} is leaky and degrades at longer delays, as shown in \cref{fig:model}D.

%%%%%%%%%%%
\input{figs_text/fig3_text.tex}
%%%%%%%%%%%

% -----------------------------------------------------------
\subsection*{Extinction and re-acquisition}
% -----------------------------------------------------------

The model can also account for the phenomenon of extinction. To investigate this, we focus on the case in which the \textit{RNN} only needs to learn a single \textit{CS-US} pair in the delay conditioning task described before. We keep the same trial structure, except that the \textit{US} is not shown at all, and the trial duration is extended (\cref{fig:extinction}A). The latter is important because in extinction, the computation of surprise in equation \ref{eqn:surprise} is triggered $t_\text{wait}$ seconds after the normal time the \textit{US} would appear, where $t_\text{wait}$ is the time after which the \textit{US} is no longer expected. Without loss of generality, we set $t_\text{wait}=5$ seconds.

As shown in \cref{fig:extinction}B, the network learns this association with a small number of trials. At this point the extinction regime is introduced by presenting the same \textit{CS} in isolation, and as a result the learned association rapidly disappears from the network (\cref{fig:extinction}B,C).

\Cref{fig:extinction}D looks at the phenomenon of re-acquisition where, after a period of extinction, the same \textit{CS-US} pair is reintroduced in training. A common finding in many classical conditioning experiments is that re-acquisition is faster than the initial learning \cite{Napier1992}. To test this, we compare two cases: one in which the same \textit{US} is used during re-acquisition (shown in blue), and one in which a different \textit{US} is used during re-acquisition (shown in red). We find that re-learning an association to the same \textit{US} is faster, therefore accounting for experimental findings on re-acquisition. Furthermore, our network provides a mechanistic explanation: re-acquisition is faster because the responses of the neurons in \cref{fig:extinction}C have not decayed to zero, even though the expectation almost has. Therefore, re-learning is faster to begin with, although the new pattern catches up later.

%%%%%%%%%%%

\input{figs_text/fig4_text.tex}
%%%%%%%%%%%

% -----------------------------------------------------------
\subsection*{Phenomena arising from \textit{CS} competition}
% -----------------------------------------------------------

%%%%%%%%%%%

\input{figs_text/fig5_text.tex}
%%%%%%%%%%%

So far we have focused on experiments in which the network needs to learn one-to-one \textit{CS-US} pairings. However, some of the most interesting  findings in conditioning arise when multiple \textit{CS}s are associated with the same \textit{US}.

To explore this, we extend the model to the case in which the network can be exposed to two \textit{CS}s for each \textit{US} (\cref{fig:competition}A). Now there are two separate \textit{RNN}s of associative neurons, one for each \textit{CS}. Without loss of generality we focus on delay conditioning and therefore, for the sake of simplicity, we remove the short-term memory network and directly feed inputs for the respective \textit{CS}s (denoted by $r_\text{cs1}$ and $r_\text{cs2}$). The activity of these populations is used to decode the identity of the \textit{US}, based on the activity generated by each \textit{CS} separately. These predictions are then used to generate expectations $E_\text{cs1}$ and $E_\text{cs2}$, which denote the predicted strength generated by each of them when shown in isolation. The total expectation for the \textit{US} is then given by $E=E_\text{cs1} + E_\text{cs2}$. The same logic could be extended to more than two \textit{CS}s. For all of these experiments, we learn a single association between a pair of \textit{CS}s and a single \textit{US}, i.e. $N_\text{stim}=1$.

\Cref{fig:competition}B presents the results for a typical blocking experiment. We first present $\textit{CS}_1$ alone for the first 100 trials, resulting in the acquisition of an expectation very close to $1$. Subsequently, we start presenting both \textit{CS}s together. However, the \textit{US} is already well predicted from $\textit{CS}_1$, resulting in small surprises  after $\textit{CS}_2$ is introduced, and thus an approximate zero learning rate. Thus, in this setting the model generates the well established phenomenon of blocking.

\Cref{fig:competition}C studies an overshadowing experiment. Here we present both \textit{CS}s together from the first trial. In this case both of them develop an expectation from the \textit{US}, but neither individually reaches $1$. Instead, it is the sum of their expectations that learns the association. Thus, in this setting the model generates the well established phenomenon of overshadowing. Notice that the expectation stemming from one of the \textit{CS}s is larger than the other, which can be attributed to randomness in the weight matrix initializations.

\Cref{fig:competition}D investigates the impact of stimulus saliency in \textit{CS} competition. Salient stimuli receive more attention and generate stronger neural responses than similar but less salient ones \cite{Gottlieb1998}. We model relative saliency by multiplying the input vector $r_\text{cs1}$ of $\textit{CS}_1$, the high-saliency cue, by a constant $s_h = 1.2 $, while keeping $r_\text{cs2}$ the same. Otherwise, the task is identical to the case of overshadowing. Consistent with animal experiments, \cref{fig:competition}D shows that the more salient $\textit{CS}_1$ acquires a substantially stronger association with the \textit{US} than the less salient $\textit{CS}_2$. This results from the fact that the more salient stimulus leads to higher firing rates, and thus to stronger pre-synaptic potentials which strengthen learning at those synapses.

Finally, \cref{fig:competition}E presents the results for a typical overexpectation experiment. Here $\textit{CS}_1$ is presented alone for the first 100 trials, $\textit{CS}_2$ is then presented alone for the next 100, and starting from trial 200, both \textit{CS}s are presented together. Since at this point the \textit{CSs} already have expectations very close to $1$, their joint expectation greatly surpasses $1$. As a result, surprise is now negative, leading to unlearning of both conditioned responses, up to the point where $E_\text{cs1}+E_\text{cs2} \approx 1$.

% -----------------------------------------------------------
\subsection*{Contingency and unconditional support}
% -----------------------------------------------------------

%%%%%%%%%%%

\input{figs_text/fig6_text.tex}
%%%%%%%%%%%

So far we have considered experiments that depend on the temporal contiguity of the \textit{CS} and \textit{US}. 
Another important variable affecting conditioning is contingency; i.e., the probability with which the \textit{CS} and the \textit{US} are presented together \cite{Rescorla1968}.

To vary the level of contingency, the \textit{US} is shown in every trial, but the \textit{CS}s are presented only with some probability, which we vary across experiments. Note that this is not the only way of running contingency conditioning experiments. For example, one could change the contingency by showing the \textit{CS}s every trial and then only show the \textit{US} with some probability. This would manipulate the degree of contingency, but also introduce an element of extinction, since there are some trials in which no \textit{US} follows the \textit{CS}. We favor the aforementioned experiment because it eliminates this confound.

\Cref{fig:contingency}A involves experiments with a single \textit{CS} which is shown with different probability. Consistent with the animal literature \cite{Rescorla1968}, we find that the strength and speed of learning increases with the \textit{CS-US} contingency.

\Cref{fig:contingency}B involves experiments with two independent predictive stimuli. Every trial $\textit{CS}_1$ is shown with probability 0.8 and, independently, $\textit{CS}_2$ is shown with probability 0.4.  Unsurprisingly, we find that the \textit{CS} with the highest contingency acquires the stronger predictive response. Note that the conditioned responses do not need to add up to $1$ in this setting. 

\Cref{fig:contingency}C,D involves a different probabilistic structure for the \textit{CS}s. $\textit{CS}_1$ is shown every trial with probability 0.8, as in the previous case. But now $\textit{CS}_2$ is only shown if $\textit{CS}_1$ is present, and with various probability $P(\textit{CS}_2 \vert \textit{CS}_1)$. When $P(\textit{CS}_2 \vert \textit{CS}_1) = 0.5$, the unconditional probabilities of the two \textit{CS}s are the same as in \cref{fig:contingency}B, but the associations learnt are different. 
After an initial acquisition phase, $E_\text{cs2}$ decays monotonically to zero. More interestingly, the same effect arises if $P(\textit{CS}_2 \vert \textit{CS}_1) = 0.875$, where $P(CS2)=0.7$:  even though the two \textit{CS}s are similarly likely, $E_\text{cs2}$ decays to zero after initially going toe-to-toe with $E_\text{cs1}$. This exemplifies the heavily non-linear behavior of this phenomenon.

To explain this finding, we need to introduce the concept of \textit{unconditional support}. A \textit{CS} has unconditional support if there are trials when it is presented by itself, which means the network has to rely on it to predict the incoming \textit{US}. In \cref{fig:contingency}B, both \textit{CS}s have unconditional support, albeit $\textit{CS}_2$'s is much lower. This explains both the noisiness in $E_\text{cs2}$, which increases each time $\textit{CS}_2$ is presented alone, and the fact that $E_\text{cs2}<E_\text{cs1}$. However, the situation drastically changes when $\textit{CS}_2$ is only presented together with $\textit{CS}_1$. Here $\textit{CS}_2$ has no unconditional support. Initially, both \textit{CS}s are conditioned, until the sum of their conditioned responses reaches $1$. At that point no more positive surprise is generated for $\textit{CS}_2$. When $\textit{CS}_1$ is presented alone, $S>0$ because $E_\text{cs1}<1$, which leads to an increase in the $E_\text{cs1}$ association. When both \textit{CS}s are presented together, the sum of their conditioned responses is now greater than $1$, and therefore $S<0$ and both conditioned responses drop. As a result, over time $E_\text{cs2}$ gradually decay to zero. This also explain why $E_\text{cs}$ takes longer to decay when $P(\textit{CS}_2 \vert \textit{CS}_1)$ is high. 

In this task, $\textit{CS}_2$ is a spurious predictor of the $\textit{US}$, since it only appears if $\textit{CS}_1$ is shown, and has no additional predictive value conditional on $\textit{CS}_1$, as shown in \cref{fig:contingency}E.  Essentially, the network learns to retain the predictive relationship but erase the spurious one. Importantly, we did nothing that would bias the network towards developing this strikingly non-linear effect.

A common fallacy of causal reasoning is known as the \textit{post hoc ergo propter hoc} fallacy \cite{Hamblin1970}. It posits that the temporal proximity of two events is sufficient to infer that the earlier event is a contributing cause of the latter. This can lead to erroneous conclusions, when such temporal proximity is coincidental. In \cref{fig:contingency}C-E, $\textit{CS}_1$ is predictive of both $\textit{CS}_2$ and the \textit{US}, but $\textit{CS}_2$ is not predictive of the \textit{US}, despite it preceding it temporally. Therefore, the network can recognize the lack of predictive ability (or unconditional support) of $\textit{CS}_2$, resolving the \textit{post hoc} fallacy in this simpler predictive setting. Similar mechanisms might allow the brain to perform more advanced forms of causal reasoning.

Finally, note that compared to other conditioning phenomena, the network takes substantially longer to learn the predictive structure of the task. Combined with the fact that real world data are scarce and often ambiguous, this might explain why such fallacies often persist.

%% file: figs_text/fig1_text.tex
\begin{figure*}[t!]
  \centering
  \includegraphics[width=0.8\textwidth]{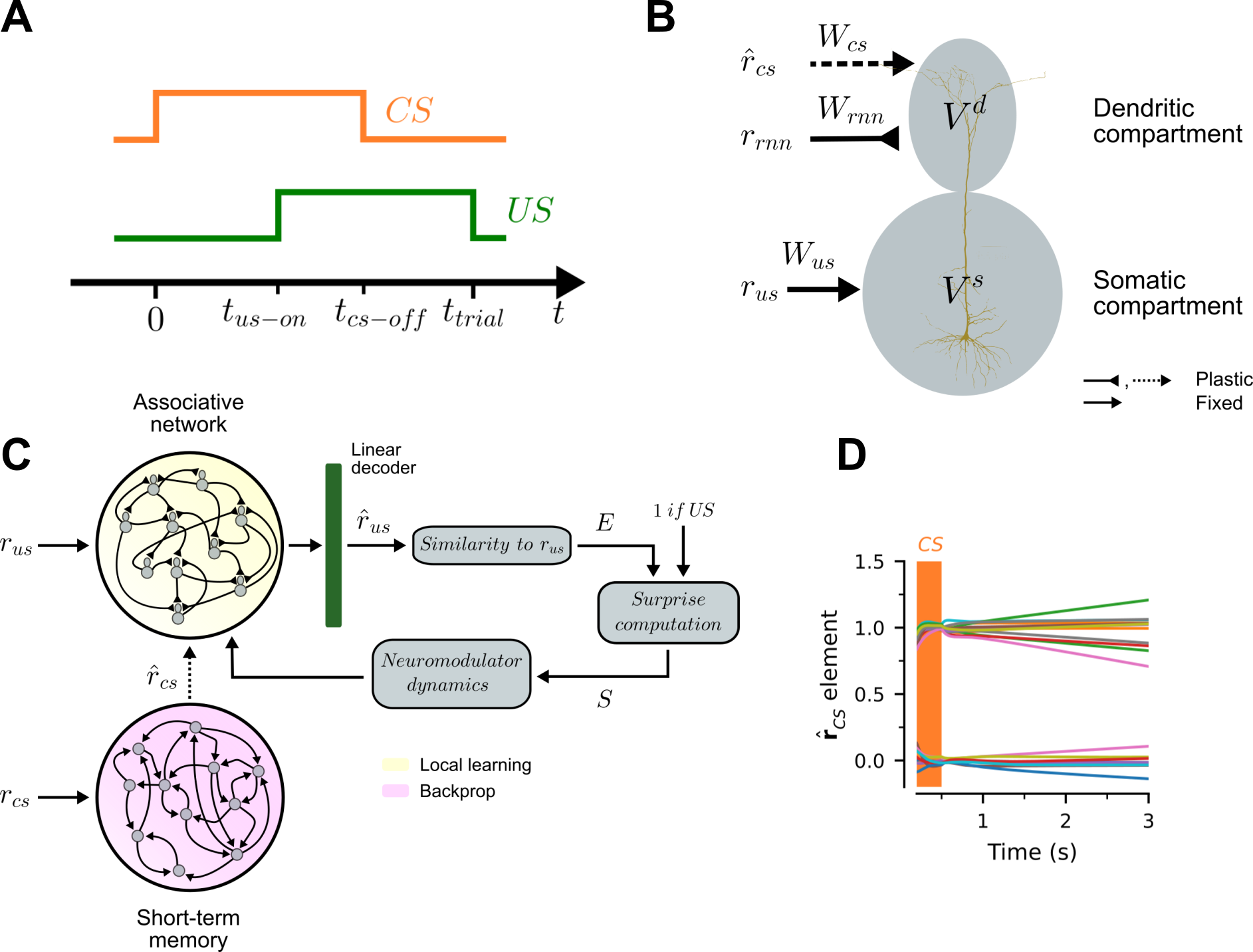}
  \caption{ \textbf{Model}. 
  (A) Every trial has a duration of $t_\text{trial}$ seconds. Trials start with the presentation of a \textit{CS}, which disappears after time $t_\text{cs-off}$. The associated \textit{US} appears at time $t_\text{us-on}$ and stays until the end of the trial. The network has to learn $N_\text{stim}$ unique \textit{CS-US} pairs. 
  (B) Associative neurons are modeled as an abstraction of a layer-5 cortical pyramidal neuron. $V^\text{s}$ and $V^\text{d}$ denote the voltage in the somatic and dendritic compartments. The somatic compartment receives as input a Boolean vector $r_\text{us}$ representing the \textit{US}. The dendritic compartment receives as inputs a vector $\hat{r}_\text{cs}$ with a short-term memory representation of the \textit{CS}, as well as recursive activity from all other neurons in the \textit{RNN}. The matrices $W_\text{rnn}$, $W_\text{cs}$ and $W_\text{us}$ denote the synaptic weights for the inputs. $W_\text{us}$ is fixed throughout the experiment. $W_\text{rnn}$ and $W_\text{cs}$ are updated over trials with training.
  (C) Full outline of the model. The associative network is made of $N_\text{rnn}$ associative neurons. The \textit{US} is presented directly to the associative neurons, whereas the \textit{CS} is presented to a short-term memory circuit that produces the short-term memory representation $\hat{r}_\text{cs}$. Learning in the associated network is gated by a surprise signal which measures the extent to which the \textit{US}, or its absence, was anticipated. The surprise signal is computed in three steps. First, throughout the trial a linear decoder is used to obtain an estimate $\hat{r}_\text{us}$ of the \textit{US} from the population vector of the associative network, denoted by $r_\text{rnn}$. Second, an expectation $E^i$ is formed according for each \textit{US} based on the similarity between $r^i_\text{us}$ and $\hat{r}_\text{us}$. These expectations determine the level of surprise $S$ associated with the arrival or absence of the \textit{US}, which then gives rise to neuromodulator dynamics that gate learning in the associative network. 
  (D) Performance of the short-term memory network in a single trial when \textit{CS}s are presented only for 500 ms. We plot the output of the memory network for several seconds. Each color denotes a different element in $r_\text{cs}$.
  }
  \label{fig:model}
\end{figure*}

%% file: figs_text/fig2_text.tex
\begin{figure*}[hbt!]
  \centering
  \includegraphics[width=0.9\textwidth]{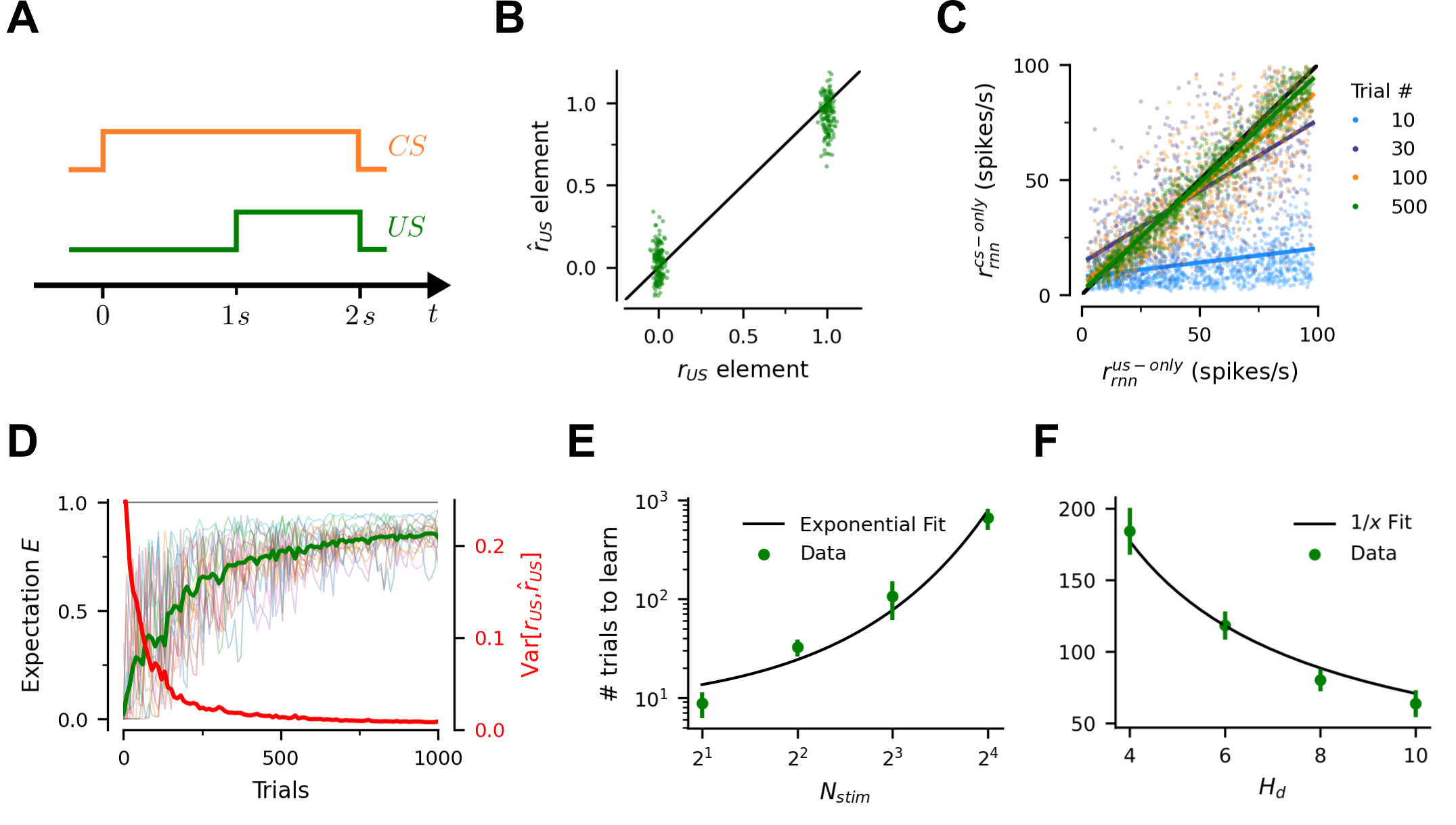}
  \caption{ \textbf{Delay conditioning and stimulus substitution}. 
  (A) Trial structure. The network is presented with $N_\text{stim}=16$ different \textit{CS-US} pairs, randomly selected in each trial. 
  (B) The network learns all of the \textit{CS-US} pairs after 500 training trials ($\approx$ 32 per pair). $r_\text{us}$ denotes the individual components of the Boolean vectors encoding each of the \textit{US}s. $\hat{r}_\text{us}$ denotes the individual components of the decoded \textit{US}s, based only on the presentation of the associated \textit{CS}s, and measured just before the \textit{US} appears. 
  (C) Evolution of population responses during learning. Colors denote trial number. Each point compares the firing rate of an associate neuron at that stage of learning for a specific \textit{CS-US} pair when only the \textit{US}, or only the associated \textit{CS} are presented. The colored lines are linear regression fits at each stage of learning. 
  (D) Evolution of predicted \textit{US} during learning. Green curve depicts the average expectation across \textit{US}s after the network is presented only with the associated \textit{CS}. Red curve depicts the distance between the true representation of the \textit{US}s ($r_\text{us}$) and their decoded representation $\hat{r}_\text{us}$ when presented only with the associated \textit{CS}. Individual pairs are shown in faint thin lines.
  (E) Number of trials required for the network to reach 80\% performance for all pairs (defined as the first time at which the average expectation $E$ across pairs exceeds $0.8$) for different numbers of stimulus pairs. Performance is measured just before the \textit{US} appears. Error bands denote $\mp$ SD computed across 5 different runs of the experiment. 
  (F) Number of trials required to reach 80\% performance for all pairs for different levels of similarity in the encoding of the \textit{CS} and \textit{US} input vectors. Error bands denote $\mp$ SD computed across 10 different runs of the experiment. 
 }
  \label{fig:delay_cond}
\end{figure*}

%% file: figs_text/fig3_text.tex
\begin{figure*}[t!]
  \centering
  \includegraphics[width=0.7\textwidth]{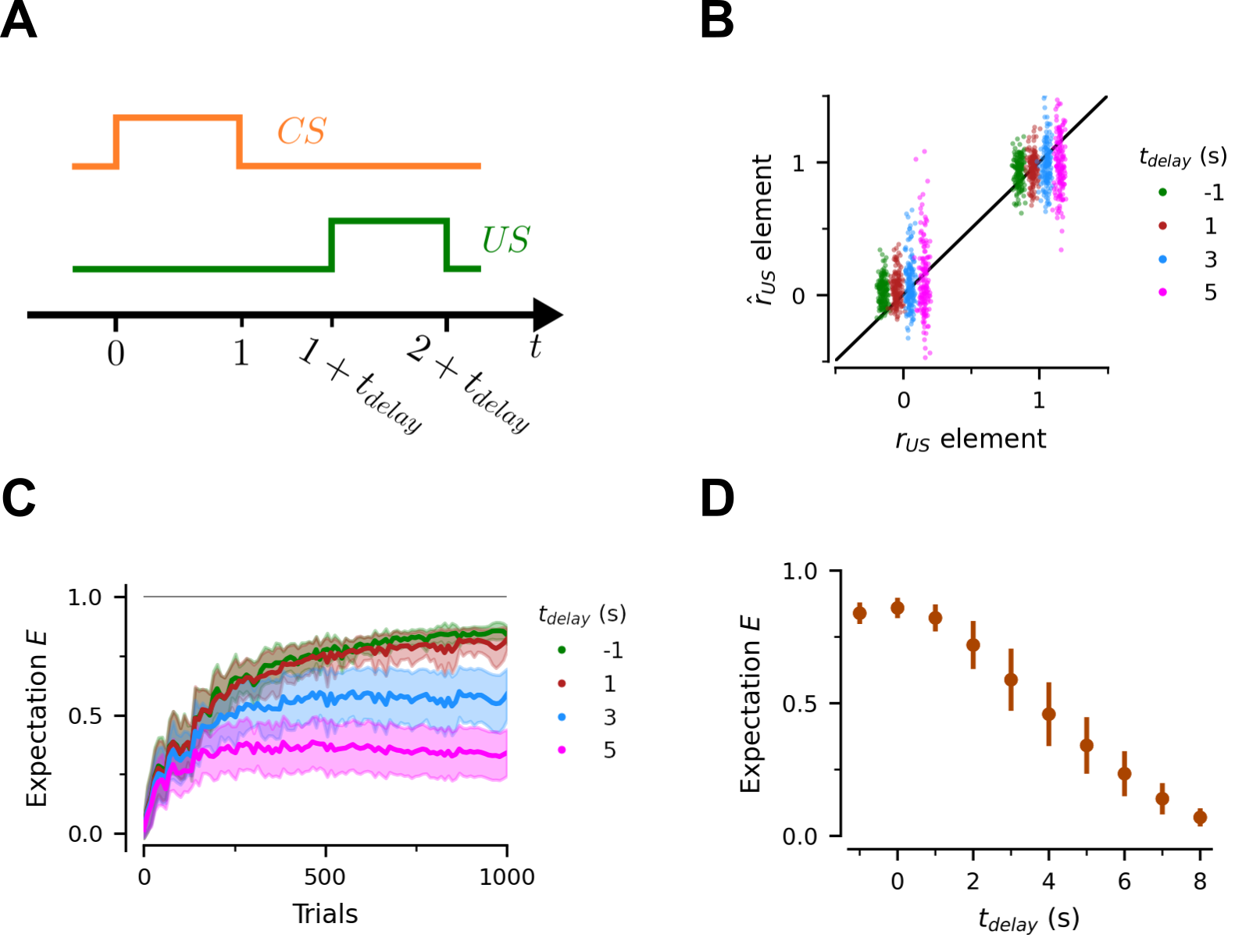}
  \caption{ \textbf{Trace conditioning}. 
  (A) Trial structure. The network is presented with $N_\text{stim}=16$ different \textit{CS-US} pairs, randomly selected in each trial. 
  (B) After 500 training trials ($\sim 32$ per pair), the network learns all of the \textit{CS-US} pairs for short $t_\text{delay}$, but struggles for longer delays. $r_\text{us}$ denotes the individual components of the Boolean vectors encoding each of the \textit{US}s. $\hat{r}_\text{us}$ denotes the individual components of the decoded \textit{US}s, based only on the presentation of the associated \textit{CS}s. For comparison purposes, we also show results for delay conditioning ($t_\text{delay} = -1$)
  (C)  Evolution of predicted \textit{US} during learning. Each curve depicts the expectation for each \textit{US} after the network is presented only with the associated \textit{CS}. Line is the mean across all stimulus pairs. Bands represent the $\mp$ SD across stimulus pairs.
  (D) Network learning performance after 500 training trials for different \textit{CS-US} delays. Bars denoted $\mp$ SD across stimulus pairs.
 }
  \label{fig:trace_cond}
\end{figure*}

%% file: figs_text/fig4_text.tex
\begin{figure*}[t!]
  \centering
  \includegraphics[width=0.6\textwidth]{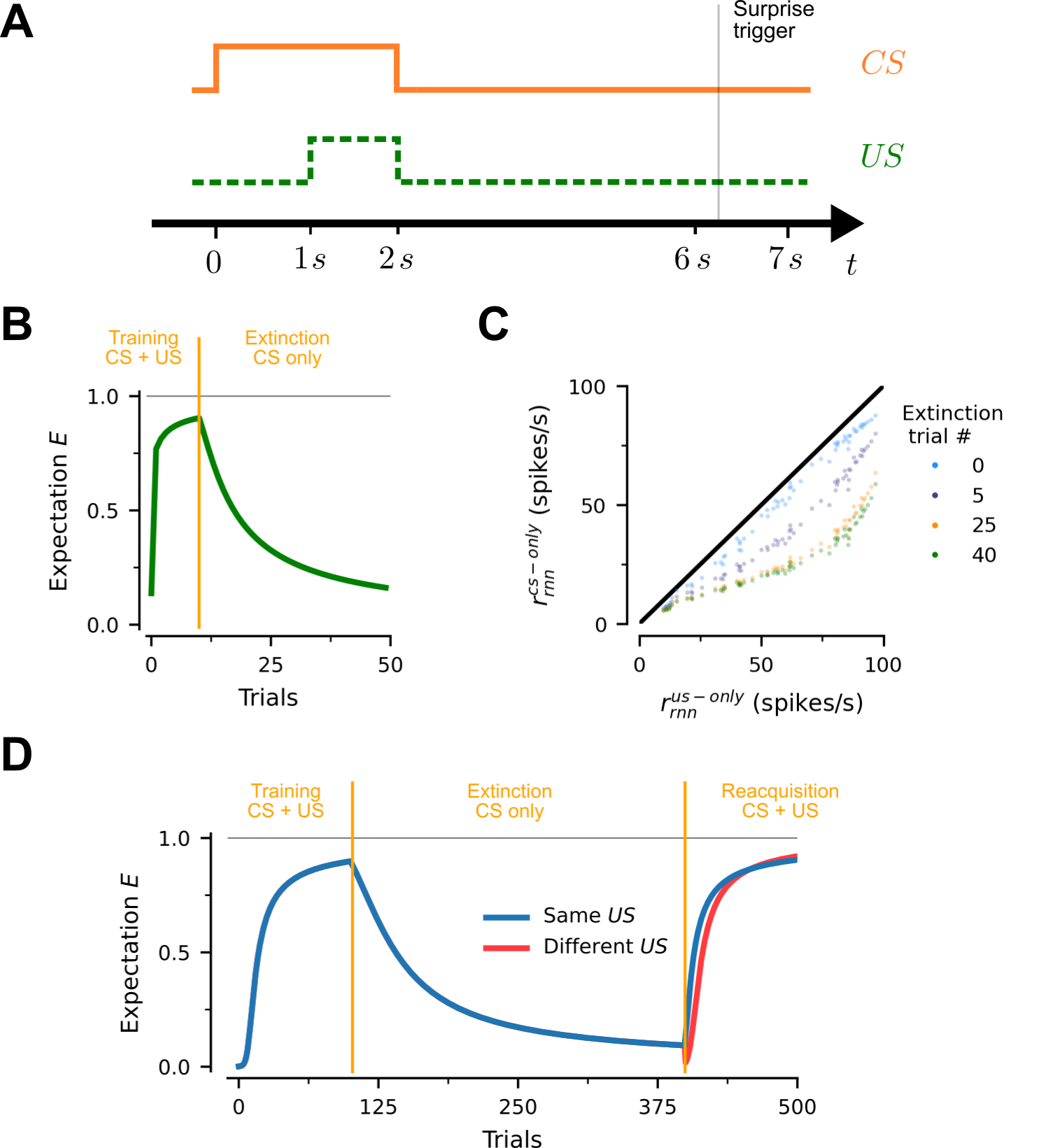}
  \caption{ \textbf{Extinction and re-acquisition}. 
  (A) Trial structure. In trials where there \textit{US} is not shown, surprise is computed at $t\approx 6$ seconds. 
  (B) Learning and extinction path for the acquisition of a single \textit{CS-US} pair. 
  (C) Evolution of population responses during extinction. Colors denote extinction trial number. Each point compares the firing rate of an associate neuron at that stage of learning for a specific \textit{CS-US} pair when only the \textit{US}, or only the associated \textit{CS} are presented. 
  (D) Learning, extinction and re-acquisition path. Blue line involves an experiment in which the same \textit{CS-US} pair is used in training and re-acquisition. Red line involves an experiment in which a new \textit{US} is used at the re-acquisition phase.
 }
  \label{fig:extinction}
\end{figure*}

%% file: figs_text/fig5_text.tex
\begin{figure*}[t!]
  \centering
  \includegraphics[width=0.9\textwidth]{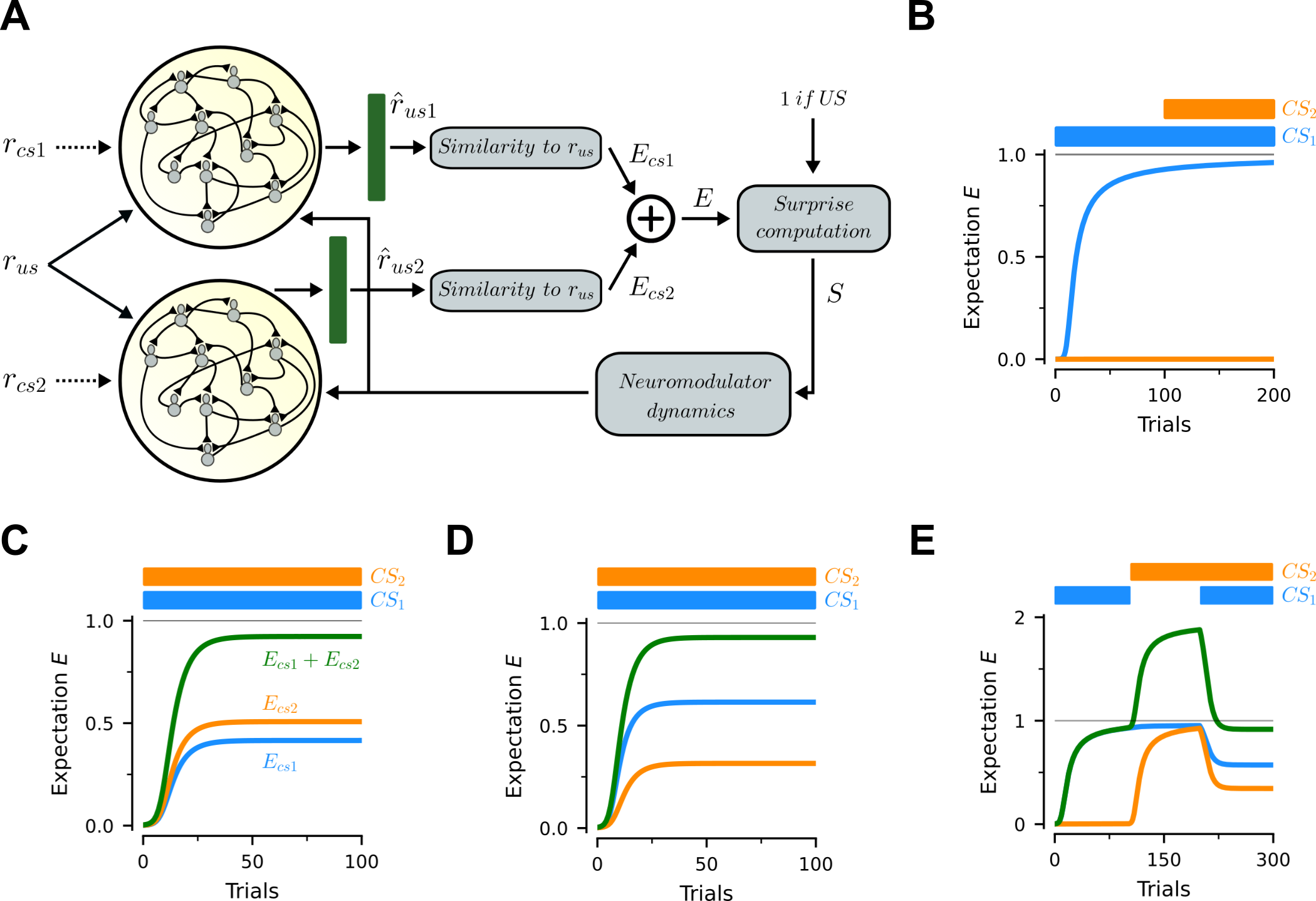}
  \caption{ \textbf{Blocking, overshadowing, saliency and overexpectation}. 
  (A) Model extension to allow for simultaneous presentation of two \textit{CS}s. Associations for $\textit{CS}_1$ and $\textit{CS}_2$ are represented in separate populations of associative neurons. The activity of each population is used to separately decode the \textit{US} and to construct expectations $E_\text{cs1}$ and $E_\text{cs2}$. The overall expectation generated by the two \textit{CS}s is given by $E=E_\text{cs1}+E_\text{cs2}$. Experiments assume that a single association between the \textit{US} and both \textit{CS}s has to be learnt. $E_\text{cs1}$ is the prediction generated by $\textit{CS}_1$ alone. $E_\text{cs2}$ is the prediction generated by $\textit{CS}_2$ alone. and $E_\text{cs1} + E_\text{cs2}$ is the prediction generated by both cues together. Since the \textit{CSs} are present throughout the trial, we omit the short-term memory networks from this exercise.
  (B) Blocking: $\textit{CS}_1$ is presented in isolation and fully learns to predict the \textit{US} before $\textit{CS}_2$ is introduced. In this case, $\textit{CS}_2$ is blocked from learning to predict the \textit{US}. 
  (C) Overshadowing: Both \textit{CS}s are presented from onset and none of them reaches the same conditioning level as when it was presented alone; instead, the sum $E$ of their expectations learns the full association. 
  (D) Saliency effects: similar to (C), but now the relative salience of $\textit{CS}_1$ has been increased by scaling up its input vector. As a result, the final conditioning level of $\textit{CS}_1$ is consistently higher than the one for $\textit{CS}_2$. 
  (D) Overexpectation: $\textit{CS}_1$ and $\textit{CS}_2$ are conditioned separately. When presented together, $E$ exceeds $1$, which leads to a negative learning rate and unlearning.}
  \label{fig:competition}
\end{figure*}

%% file: figs_text/fig6_text.tex
\begin{figure*}[t!]
  \centering
  \includegraphics[width=0.7\textwidth]{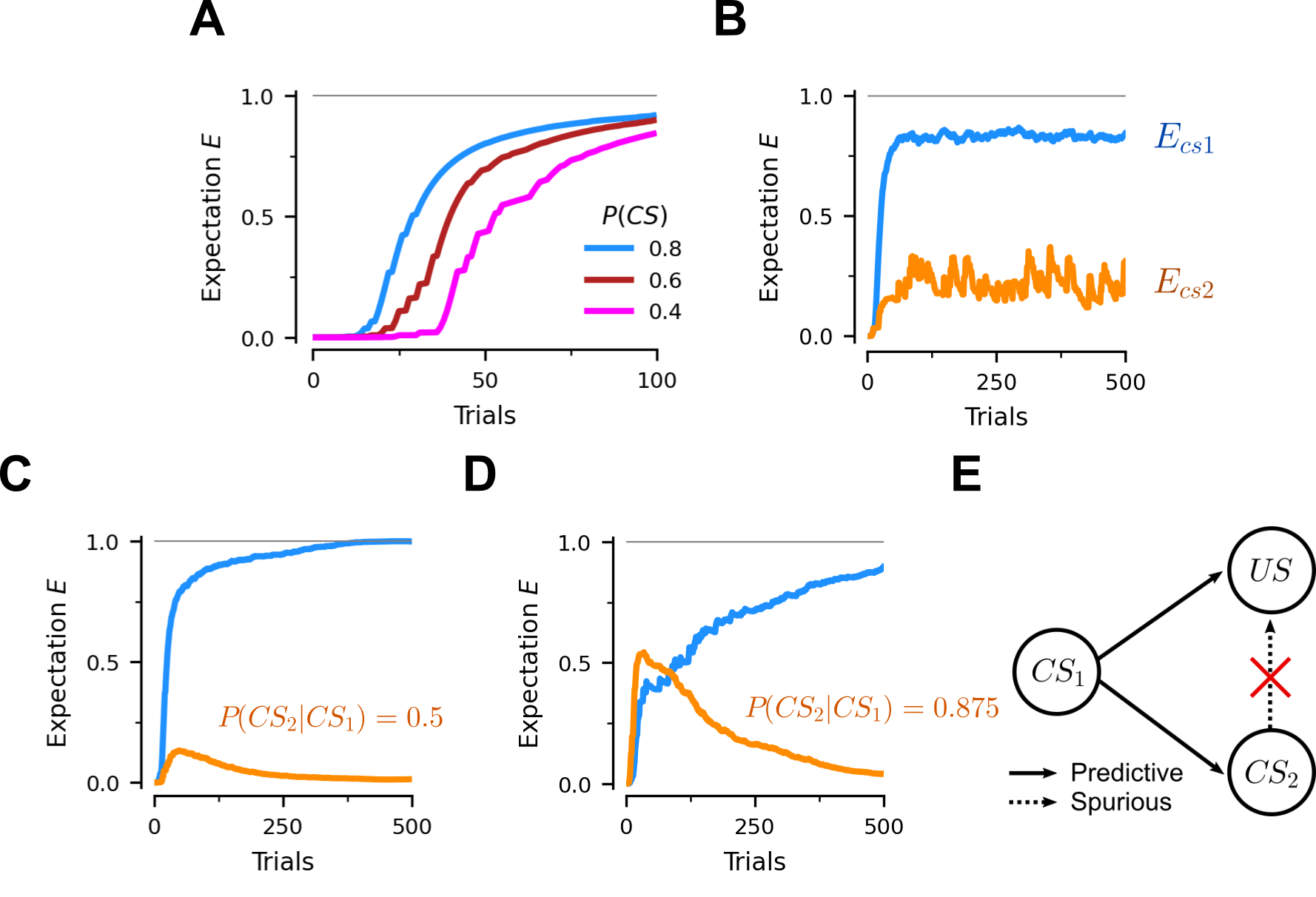}
  \caption{ \textbf{Contingency and causality}.
  The \textit{US} is shown every trial, while the contingency of the \textit{CS}s is varied. 
  (A) Impact of changing the probability of showing the \textit{CS} in every trial. Each line depicts the learning path for a different experiment. 
  (B) Experiment with two independent predictive stimuli. In every trial, $\textit{CS}_1$ is shown with probability 0.8 and $\textit{CS}_2$ is shown with probability 0.4. Blue curve is the expectation acquired by $\textit{CS}_1$ when shown by itself. Orange curve is the expectation acquired by $\textit{CS}_2$ when shown by itself.
  (C,D) Experiments with a conditional \textit{CS} structure. Every trial $\textit{CS}_1$ is shown with probability 0.8 and $\textit{CS}_2$ is shown only if $\textit{CS}_1$ is also present, with probability $P(\textit{CS}_2 \vert \textit{CS}_1)$.
  (E) The network learns to ignore spurious predictors. Since $\textit{CS}_2$ is conditionally dependent on $\textit{CS}_1$, our network gradually phases out any explanatory power of $\textit{CS}_2$, as more evidence that the $\textit{US}$ is never caused by the $\textit{CS}_2$ by itself arrives.
 }
  \label{fig:contingency}
\end{figure*}

%% file: discussion.tex
%!TEX root = main.tex

\section*{Discussion}

%Summary
The ability to engage in stimulus-stimulus associative learning provides a crucial evolutionary advantage. The cerebral cortex might contribute to this evolutionary edge by exploiting representational \cite{Rigotti2013} and architectural \cite{Larkum1999} inductive biases present in the cortical microcircuit \cite{Nieuwenhuys1994}. We here propose a recurrent neuronal network model of how the cortex can implement stimulus substitution, which allows the same set of neurons to encode multiple stimulus-stimulus associations. The model relies on the properties of two-compartment layer-5 pyramidal neurons, which based on recent experimental findings, we refer to as associative neurons. These neurons can act as coincidence detectors for information about the \textit{US} arriving at their somatic compartment and information about the \textit{CS} arriving at their dendritic compartment \cite{Larkum1999,Larkum2013,Doron2020}. Coincidence detection allows for a biologically plausible synaptic plasticity rule that, after learning, results in neurons that would normally fire in the presence of the \textit{US} to respond in the same manner when the \textit{CS} is presented. At the population level, this means that the pattern of neural activity corresponding to the \textit{CS} can be morphed into the one corresponding to the \textit{US}, leading to stimulus substitution. 

Our model accounts for many of most important conditioning phenomena observed in animal experiments, including delay conditioning, trace conditioning, extinction, blocking, overshadowing, saliency effects, overexpectation and contingency effects. The model is able to learn multiple \textit{CS-US} associations with a degree of training that is commensurate with animal experiments. Significantly, the model performs well across a wide variety of conditioning tasks without experiments-specific parameter fine-tuning.

We also show that some influential models of three-factor Hebbian learning rules - Oja's rule \cite{Oja1982} and the BCM rule \cite{Bienenstock1982} - fail to learn generic stimulus-stimulus associations due to their unsupervised nature. Hebbian rules have demonstrable autoassociative \cite{Hopfield1982} and heteroassociative \cite{Sompolinsky1986} capabilities, and when augmented with eligibility traces they have been shown to account for neuronal-level reinforcement learning \cite{Urbanczik2009,Vasilaki2009,Fremaux2010}. Still, they struggle with pattern-to-pattern associations when representations are mixed. This is because Hebbian rules are purely unsupervised, and therefore provide no guarantee that the impact of the \textit{CS} will be eventually shaped to be identical to the one of the \textit{US}. Instead, network performance heavily depends on implementation details, like training history, task details and stimulus statistics. As a result, decoding from a population encoding several associations is hampered by the fact that activation levels for individual neurons when exposed to the \textit{CS} will more often than not be off from those resulting from exposure to the corresponding \textit{US}.

% Lit

% TBD: ADD Here discussion on how literature uses implausible learning rules?

Related work utilized a predictive learning rule similar to the one used here to account for prospective coding of anticipated stimuli \cite{Brea2016}. While prospective coding might also be involved in conditioning, their study differs in several ways. First, their learning rule is timing-dependent; it succeeds in a delayed pair associative learning task, but it would require re-learning when the relative timing of the \textit{US} in relation to the \textit{CS} is variable. In contrast, our learning rule applies to arbitrary task timings. Second, their learning rule lacks gating which, unless strict conditions are met (dendritic and somatic activity conditioned on a stationary Markov chain), leads to reduced responses and even catastrophic forgetting. Furthermore, adding gating is not feasible in their model, because learning needs to bootstrap before the presentation of the delayed stimulus, and gating would inactivate learning at these times.

% About mdoel mechs
Several features of the model are worth emphasizing. 

First, the proposed \textit{RNN} leverages architectural inductive biases in the form of two-compartment associative neurons. These associative neurons are the most common neuron type in the mammalian cortex \cite{Nieuwenhuys1994}. This is likely no coincidence; once evolution stumbled upon their usefulness in predicting external contingencies, it might have favored them. While subcortical \cite{Christian2003} and even single-neuron \cite{Gershman2021} mechanisms for conditioning exist, the mechanism that we propose can handle mixed representations, and thus allow animals with a cerebral cortex to flexibly learn large numbers of associations. 

The structure of the associative neuron is ideal for stimulus-stimulus learning. Feedforward inputs, like the \textit{US} representations, arrive near the soma in layer-5 and directly control the neuron's firing rate. Feedback inputs, like the \textit{CS} representations and the activity of other cortical neurons, arrive at the distal dentrites in layer-1 \cite{Larkum2013}. This compartmentalized structure allows the signals to travel independently, and get associated via a cellular mechanism known as BAC firing \cite{Larkum1999}. Specifically, it has been shown that these cells implement coincidence detection, whereby feedforward inputs trigger a spike which backpropagates to the distal dendrites and concurrently feedback input arrives at these dendrites, then plateau calcium potentials are initiated in the dendritic compartment \cite{Larkum1999}. These plateau potentials result in the neuron spiking multiple times subsequently and learning occurs in the distal dendrites, so that feedback inputs can elicit spikes alone in the future, without the need for external information.

Second, a prerequisite for the biological plausiblity of the learning rule used in the model is that backpropagating action potentials to be disentangled from postsynaptic potentials at the dendritic compartment. Only then can the two critical components in our learning rule, $f(V^\text{s})$ and $f(p' V^\text{d})$ in eq. \ref{eqn:learn_rule_main} be compared. Since backpropagating action potentials (denoted by $f(V^\text{s})$ in the model) do not need to travel far, they experience minimal  attenuation \cite{Larkum1999} and therefore they maintain some of their high-frequency components, which could be used at synapses to differentiate them from slower postsynaptic potentials (denoted by $V^\text{d}$ in the model). As a result, only a static transformation of this last term is needed to compare the two signals. Consequently, the learning rule relies only on information locally available at each synapse, which is a prerequisite for biological plausibility.

Third, our model suggests multiple functional roles for gating. It limits learning to episodes that appear to have behavioral significance. Gating also prevents drifting of learned associations due to a lack of perfect self-consistency between $f(V^\text{s})$ and $f(p' V^\text{d})$ in the learning rule \cite{Urbanczik2009}, which is expected in a biological system subject to noise and approximate computation. In addition, gating provides a critical global reference signal when multiple \textit{CS}s  are available at the same time.

% TBD/potential. Fifth, Add link w/ predictive coding.

%Limitations
The model also has some limitations to be addressed in future work. Most importantly, it does not account for spontaneous recovery of previously learnt associations after extinction. In our model, 
extinction stems from the decay of the response of the associate neurons to the \textit{CS}, a mechanism akin to unlearning, which erases previous learning, and thus does not allow for spontaneous recovery or faster re-acquistion. The extinction mechanism proposed here is complementary to inhibitory learning, the mechanism initially put forth by Pavlov to explain spontaneous recovery.

%TBD: fix if model leads to faster reacquisiion. Instead, it accounts more permanent erasure of associations, although since the final \textit{CS}-induced pattern still somewhat resembles the \textit{US}-induced (with reduced gain), re-acquisition will be faster. This is because forgetting is also contingent on expectation, and once the \textit{CS} is no longer close enough to the \textit{US} to result in $E>0$, unlearning will cease and remnants of the association will remain. .

In the case of experiments with multiple \textit{CS}s, the model assumes that different neuronal population implements separate \textit{RNN}s to learn the associations for each of them. Although the two populations interact indirectly through the surprise signals, they each learn to predict the \textit{US} on their own. The existence of separate populations might be justifiable when the \textit{CS}s involve different sensory modalities (e.g., sound and vision), or very different spatial locations, but not necessarily when they are presented simultaneously. Extending the model to include differential routing of simultaneously presented stimuli is an open question for future work. 

Another direction for future work is to account for more psychological aspects of conditioning by developing a larger model that incorporates other forms of learning and generalization like model-based strategies also thought to take place in the PFC \cite{Wang2018}, or to allow for context-dependent computation to resolve conflicts among competing stimuli \cite{Mante2013}. In these larger models, our network would model the stimulus substitution component. 

The model allows to differentiate between conditioning effects that can be accounted by low-level, synaptic plasticity mechanisms, versus other high level explanations. At its core, the model performs stimulus substitution at the neuronal level, via a gradual acquisition process \cite{Thorndike1898,Rescorla1972,Sutton1981}. Despite that, the model is still capable of rapid, few-shot learning, especially when the number of associations is small compared to size of the network (\cref{fig:delay_cond}E). Yet, for rapid learning in more complicated scenarios, fast inference based on prior knowledge might be necessary \cite{Lake2016}.

Finally, our model suggest an alternative role for representational inductive biases in the form of mixed selectivity, other than readout flexibility \cite{Fusi2016}: it permits the efficient packing of multiple stimulus-stimulus associations within the same neuronal population, which might confer cortical animals the evolutionary edge.

%% file: methods.tex
%!TEX root = main.tex

\section*{Methods}

% -----------------------------------------------------------
\subsection*{\textit{RNN} of associative neurons}
% -----------------------------------------------------------

The central element of the model is a \textit{RNN} of $N_\text{rnn}$ associative neurons. The goal of the network is to learn to predict the identity of the upcoming \textit{US} from the presentation of the corresponding \textit{CS}, by reproducing the \textit{US} population vector when only the \textit{CS} is presented. Each associative neuron is a two-compartment rate neuron modelled after layer-5 pyramidal cortical neurons \cite{Larkum1999,Urbanczik2014}. The somatic compartment models the activity of the soma and apical dendrites of the neuron, while the dendritic compartment models the activity of distal dendrites in cortical layer-1. As depicted in \cref{fig:model}B, the somatic compartment receives $r_\text{us}(t)$ as input, whereas the dendritic compartment receives $\hat{r}_\text{cs}(t)$ as well feedback activity from the all the \textit{RNN} units, which is denoted by $r_\text{rnn}(t)$.

The instantaneous firing rate of the associative neurons is a sigmoidal function of the somatic voltage $V^\text{s}$:
\begin{equation}
r_\text{rnn} = \frac {f_\text{max}}{1+\exp\left[-\beta(V^\text{s}-V_{1/2})\right]}.
\label{eqn:act_fun}
\end{equation} 
This activation function is applied element-wise to the vector~$V^\text{s}$, which represents the instantaneous somatic voltage in each associative neuron. $f_\text{max}$ sets the maximum firing rate of the neuron, $\beta$~is the slope of the activation function, and~$V_{1/2}$ is the voltage level at which half of the maximum firing rate is attained. We set $f_\text{max}$ to a reasonable value for cortical neurons, and choose appropriate values for $\beta$ and $V_{1/2}$ so that the whole dynamic range of the activation function is used and firing rates when somatic input is present are relatively uniform. See Table 1 for a description of all model parameters, and Table S1 for their justification. 

The somatic voltages, and thus the firing rates, are determined by the following system of differential equations:

\begin{itemize}

\item The associative neurons receive an input current to their dendritic compartments, denoted by $I^\text{d}$,  which obey:
\begin{equation}
\tau_\text{s} \frac {{\textrm d} I^\text{d}}{{\textrm d}t} = - I^\text{d} + W_\text{cs}  \, \hat{r}_\text{cs} + W_\text{rnn}  \, r_\text{rnn} 
\label{eqn:cur_den}
\end{equation}
where $W_\text{rnn}$ is the matrix of synaptic weights between any pair of associative neurons (dimension: $N_\text{rnn} \times N_\text{rnn}$), $W_\text{cs}$ is the matrix of synaptic weights for the \textit{CS} input (dimension: $N_\text{rnn} \times N_\text{inp}$), and $\tau_\text{s}$ is the synaptic time constant.

\item The dynamics of the voltage in the dendritic compartments $V^\text{d}$ are given by:
\begin{equation}
\tau_\text{l} \frac {\textrm{d} V^\text{d}}{\textrm{d}t} = - V^\text{d} + I^\text{d};
\label{eqn:pot_den}
\end{equation}
i.e. it is a low-pass filtered version of the dendritic current $I^\text{d}$ with the leak time constant $\tau_\text{l}$. For simplicity, voltages and currents are dimensionless in our model. Therefore the leak resistance of the dendritic compartment is also dimensionless and set to unity.

\item The voltages of the somatic compartments, denoted by $V^\text{s}$, are given by:
\begin{equation}
C \frac {\textrm{d} V^\text{s}}{\textrm{d}t} = -g_L V^\text{s} - g_D (V^\text{s}-V^\text{d}) + I^\text{s}
\label{eqn:pot_som}
\end{equation}
where $C$ is the somatic membrane capacitance, $g_L$ is the leak conductance, $g_D$ is the conductance of the coupling from the dendritic to the somatic compartment, and $I^\text{s}$ is a vector of input currents to the somatic compartments. Note that this specification assumes that the time constant for the somatic voltage is one, or equivalently, that it is included in $C$.

\item The vector $I^\text{s}$ of input currents to the somatic compartment is given by:
\begin{equation}
I^\text{s} = g_\text{e} \odot (E_\text{e} - V^\text{s}) + g_\text{i} \odot (E_\text{i} - V^\text{s})
\label{eqn:ff_input}
\end{equation}
where $g_\text{e}$ and $g_\text{i}$ are vectors describing the time-varying excitatory and inhibitory conductances of the inputs, $E_\text{e}$ and $E_\text{i}$ are the reversal potentials for excitatory and inhibitory inputs, and $\odot$ denotes the Hadamard (element-wise) product.

\item The vectors of excitatory and inhibitory conductances $g_\text{e}$ and $g_\text{i}$ for the somatic compartment are described, respectively, by the following two equations:
\begin{equation}
\tau_\text{s} \frac {\textrm{d} g_\text{e}}{\textrm{d}t} = - g_\text{e} + \left[ W_\text{us}  \right]_+ r_\text{us}
\label{eqn:exc_cond}
\end{equation}
and
\begin{equation}
\tau_\text{s} \frac {\textrm{d} g_\text{i}}{\textrm{d}t} = - g_\text{i} + \left[- W_\text{us}  \right]_+ r_\text{us} + g_\text{inh}
\label{eqn:inh_cond}
\end{equation}
where $W_\text{us}$ is a matrix describing the synaptic weights for the \textit{US} inputs to the somatic compartments (dimension: $N_\text{rnn} \times N_\text{inp}$), $\tau_\text{s}$ is the same synaptic time constant used in equation \ref{eqn:cur_den}, $g_\text{inh}$ is a constant inhibitory conductance of all associative neurons, and $\left[ . \right]_+$ is the rectification function applied element-wise.

\end{itemize}

The model implicitly assumes zero resting potentials for the somatic and dendritic compartments. In addition, we assume that there is no input to the \textit{RNN} between trials, and that the inter-trial interval is sufficiently long so that the variables  controlling activity in the associative neurons reset to zero between trials. The differential equations describing activity within trials are simulated using the forward Euler method with time setp $\Delta t = 1$ ms.

At the beginning of the experiment, all synaptic weight matrices are randomly initialised, independently for each entry, using a normal distribution with mean 0 and standard deviation $1/\sqrt{N_\text{rnn}}$, as is standard in the literature. Note that since associative neurons are pyramidal cells, the elements of $W_\text{rnn}$ are restricted to positive values; hence we use the absolute value of those random weights. 

$W_\text{us}$ stays fixed for the entire experiment. $W_\text{rnn}$ and $W_\text{cs}$ are plastic and updated using the learning rules described next.

\input{tables/table_1.tex}

% -----------------------------------------------------------
\subsection*{Synaptic plasticity rule}
% -----------------------------------------------------------

We utilize a synaptic plasticity rule inspired by \cite{Larkum1999,Larkum2013,Doron2020}, where the firing rate of the somatic compartment in the presence of the \textit{US} acts like a target signal for learning the weights $W_\text{rnn}$ and $W_\text{cs}$  (see \cite{Urbanczik2014} for the initial spike-based learning rule, and \cite{Vafidis2022} for the rate-based formulation). The learning rule modifies these synaptic weights so that, after learning, \textit{CS} inputs can predict the responses of the \textit{RNN} to the \textit{US}s.

Consider the synaptic weights from input neuron $j$ to associative neuron $i$, for either the \textit{RNN} or the \textit{CS} inputs. The weights are updated continuously during the trial using the following rule:
\begin{equation}
\Delta W_{ij} = \eta (S) \left[\, f(V^\text{s}_i)-f(p^\prime \, V^\text{d}_i)\,\right]\, P_j 
\label{eqn:learn_rule}
\end{equation}
where $\eta (S)$ is a variable learning rate that depends on the instantaneous level of a surprise signal $S$, $p^\prime$ is an attenuation constant derived below, and $P_j$ is the postsynaptic potential in input neuron $j$.

The postsynpactic potential $P_j$ has a simple closed form solution detailed in \cite{Vafidis2022}. In particular, it is a low-passed filtered version of the neuron's firing rate, so that
\begin{equation}
P_j(t) = H(t) * r_j(t), 
\end{equation}
where $*$ denotes the convolution operator, and $H$ is the transfer function given by 
\begin{equation}
H(t)=\frac{1}{\tau_\text{l} - \tau_\text{s}} \left[ \exp(-\frac{t}{\tau_\text{l}}) - \exp(-\frac{t}{\tau_\text{s}}) \right] u(t)
\end{equation}
and $u(t)$ is the Heaviside step function that takes a value of $1$ for $t>0$ and a value of $0$ otherwise.

As noted in \cite{Vafidis2022}, for constant $\eta$ the learning rule is a predictive coding extension of the classical Hebbian rule. When $\eta$ is controlled by a surprise signal, as in our model, it can be thought of a predictive coding extension of a three-factor Hebbian rule \cite{Fremaux2010,Fremaux2016}. 

Importantly, all of the terms in the learning rule are available at the synapses in the dendritic compartment, making this a local, biologically plausible learning rule. The firing rate of the neuron $f(V^\text{s}_i)$ is available due to backpropagation of action potentials \cite{Larkum1999}.  $f(p' V_i^\text{d})$ is a constant function of the local voltage $V_i^\text{d}$  computed locally in the dendritic compartment even when the somatic input is present. By definition, postsynaptic potentials are available at the synapse.

There are a total number of $N_\text{train}$ training trials, divided among all \textit{CS-US} pairs. After each training trial we measure the state of the \textit{RNN} off-line by inputing one $r_\text{cs}$ at a time without the \textit{US}, keeping the network weights constant, and measuring the output produced by the model at that stage of the learning process.

% -----------------------------------------------------------
\subsection*{Convergence of synaptic plasticity rule}
% -----------------------------------------------------------

 To understand how and why the learning rule works, it is useful to characterize the somatic voltages, and thus their associated firing rates, in different trial conditions.

Consider first the case in which only the \textit{CS} is presented, so the associative neurons only receive dendritic input. In this case the somatic voltages converge to a steady-state given by 
\begin{equation}
V^\text{ss} = \frac {g_D}{g_D+g_L} V^\text{d} .
\label{eqn:pot_att}
\end{equation}
In other words, the somatic voltages converge simply to an attenuated level of the dendritic voltages, with the level of attenuation given by $p=\frac {g_D}{g_D+g_L}$. In this case, the firing rates of the associative neurons converge to
\begin{equation}
r^\text{cs-only}_\text{rnn} = f(V^\text{ss})
\label{eqn:fb_fr}
\end{equation}
This follows from the fact that the dendritic voltage is determined only by equations \ref{eqn:cur_den} and \ref{eqn:pot_den}, and thus is not affected by the state of the somatic compartment, and by the fact that in the absence of \textit{US} input $I^\text{s}=0$. The result then follows immediately from equation \ref{eqn:pot_som}.

Next consider the case in which only the \textit{US} is presented. In this case equations \ref{eqn:cur_den} and \ref{eqn:pot_den} imply that $V^\text{d}=0$, and it then follows from equations \ref{eqn:pot_som} and \ref{eqn:ff_input} that the steady-state somatic voltage, when $I^\text{s}=0$, is given by 
\begin{equation}
V^\text{eq}(t) = \frac{g_\text{e} E_\text{e} + g_\text{i} E_\text{i}}{g_\text{e} + g_\text{i}}
\label{eqn:eq_volt}
\end{equation}
and that the firing rates of the associative  neurons become
\begin{equation}
r^\text{us-only}_\text{rnn} = f(V^\text{eq}).
\label{eqn:ff_fr}
\end{equation}

Finally consider the case in which the associative neurons receive input from both the \textit{CS} and the \textit{US}. We follow \cite{Brea2016} to derive the steady-state solution for the somatic voltage in this case. Provided inputs to the circuit, which are in behavioral timescales, change slower than the membrane time constant ($C/g_L= 20 \, \textrm{ms}$), equation \ref{eqn:pot_som} reaches a steady-state given by
\begin{equation}
V^\text{s}(t) \approx \kappa V^\text{ss} + (1-\kappa) V^\text{eq},
\label{eqn:ss_volt}
\end{equation}
where $\kappa(t) = \frac{g_D+g_L}{g_D+g_L+g_\text{e}+g_\text{i}} \in (0,1]$ performs a linear interpolation between the steady-state levels reached where only the \textit{CS} or the \textit{US} are presented.

Practically, when there is no \textit{US}-input, $V^\text{ss}$ slightly precedes $V^\text{s}$ due to the non-zero dendritic-to-somatic coupling delays, resulting in slight overestimation of the firing rate upon \textit{CS} presentation. This can be accounted for by introducing an additional small attenuation, so that $p^\prime = a \frac{g_D}{g_D + g_L} = a p$ in equation \ref{eqn:learn_rule}, with $a=0.95$.

Learning is driven by a comparison of the firing rates of the associative neurons in the presence of both the \textit{CS} and the \textit{US}, and the firing rates if they only receive input from the \textit{CS}. Importantly, this can happen online and without the need for separate learning phases, because an estimate of the latter can be formed in the dendritic compartment at all times. Learning is achieved by modifying $W_\text{rnn}$ and $W_\text{cs}$ to minimize this difference. We can use the expressions derived in the previous paragraphs to see why the synaptic learning rule converges to synaptic weights for which $r^\text{cs-only}_\text{rnn} = r^\text{both}_\text{rnn}$.

Take the case in which associative neurons underestimate the activity generated by the \textit{US} inputs when exposed only to the \textit{CS} (i.e. $V^\text{ss} < V^\text{eq}$). In this case,  $V^\text{ss} < V^\text{s} < V^\text{eq}$ and $I^\text{s}>0$. Then from equation \ref{eqn:learn_rule} we find that $\Delta w>0$, leading to a futures increase in associative neuron activity in response to the \textit{CS}.

The same logic applies in opposite case, where the associative neurons overestimate the activity generated by the \textit{US} inputs when exposed only to the \textit{CS}. In this case,  $V^\text{ss} > V^\text{s} > V^\text{eq}$ and $I^\text{s}<0$, which leads to a future decrease in associative neuron activity in response to the \textit{CS}.

Given enough training, this leads to a state where $V^\text{ss} \approx V^\text{eq}$ and at which learning stops ($\Delta w \approx 0$). When this happens, we have that
\begin{equation}
r^\text{cs-only}_\text{rnn} = f(V^\text{ss}) \approx f(V^\text{eq})=r^\text{both}_\text{rnn},
\label{eqn:ff_fr_learn}
\end{equation}
so that the \textit{RNN} responses to the \textit{CS} become fully predictive of the activity generated by the \textit{US}, when presented by themselves.

% -----------------------------------------------------------
\subsection*{\textit{US} decoding}
% -----------------------------------------------------------

Up to this point the model has been faithful to the biophysics of the brain. The next part of the model is designed to capture the variable learning rate $\eta$ in equation \ref{eqn:learn_rule}, and thus is more conceptual in nature. Our goal here is simply to provide a plausible model of the factors affecting the learning rates for the \textit{RNN}. As illustrated in \cref{fig:model}C, this part of the model involves three distinct computations: decoding the \textit{US} from the \textit{RNN} activity, computing expectations about upcoming \textit{US}s, and computing the surprise signal $S$.

The brain must have a way to decode the upcoming \textit{US}, or its presence, from the population activity in the \textit{RNN} at any point during the trial. This prediction is represented by the time-dependent vector $\hat{r}_\text{us}(t)$. For the purposes of our model, we will use the optimal linear decoder $D$ (dimension: $N_\text{rnn} \times N_\text{inp}$), so that
\begin{equation}
\hat{r}_\text{us}(t) = r_\text{rnn}(t)^\intercal D.
\end{equation}

The optimal linear decoder $D$ is constructed as follows. First, for each \textit{US} $i=1,...,N_\text{stim}$ define the row vector $\phi_i$ describing the steady-state firing rate the each associative neuron that arises when it is presented alone. Then define an activity matrix $\Phi$ by stacking vertically these $N_\text{stim}$ row vectors (dimension: $N_\text{stim} \times N_\text{rnn}$). $\Phi$ is built using the initial random weights $W_\text{rnn}$, before learning has taken place. Second, define a target matrix $T$ (dimension:  $N_\text{stim} \times N_\text{inp}$) to be the row-wise concatenated set of \textit{US} input vectors $r_\text{us}$. Then, if $D$ perfectly decodes the \textit{US} from the \textit{RNN} activity, when only the \textit{US}s are presented, we must have that
\begin{equation}
\Phi D = T .
\end{equation}
It then follows that 
\begin{equation}
D = \Phi^+ T ,
\label{eqn:dec_est}
\end{equation}
where $^+$ denotes the Moore-Penrose matrix inverse. A desirable property of the Moore-Penrose inverse is that if equation \ref{eqn:dec_est} has more than one solutions, it provides the minimum norm solution, which results in the smoothest possible decoding. 

Note that the decoder, which could be implemented in any downstream brain area requiring information about \textit{US}s, is completely independent of the input representations of the \textit{CS}s. Instead, it is determined before learning given only knowledge of the \text{US}s, and is kept fixed throughout training.

% -----------------------------------------------------------
\subsection*{\textit{US} expectation estimation}
% -----------------------------------------------------------

Since the \textit{US}s are primary reinforcers, it is reasonable to assume that their representations, $r^i_\text{us}$ for $i=1,...,N_\text{stim}$ , are stored somewhere in the brain. Then an expectation for each \textit{US} can be formed by
\begin{equation}
E^i(t) = \exp (-\kappa \| \hat{r}_\text{us}(t) - r^i_\text{us} \|^2 ) ,
\label{eqn:R_est}
\end{equation}
where $\| \hat{r}_\text{us}(t) - r^i_\text{us} \|$ is Euclidean distance between the stored and the decoded representations for each \textit{US} at time $t$, and $\kappa$ controls the steepness of the Gaussian kernel. Recognizing that the ability to discriminate these patterns increases with the Hamming distance $H_\text{d}$, we set the precision to be inversely proportional to $H_\text{d}$ i.e. $\kappa = \left ( \frac{8}{H_\text{d}} \right )^2$.

Note that $E^i$ takes values between 0 and 1, and equals 1 only when the \textit{US} is perfectly decoded (i.e., when $\hat{r}_\text{us} = r^i_\text{us}$). Thus, $E^i$ can be interpreted as a probabilistic estimate for each \textit{US} that is computed  throughout the trial. To simplify the notation, we denote the expectation for the \textit{US} associated with the trial as $E$.

% -----------------------------------------------------------
\subsection*{Surprise based learning rates}
% -----------------------------------------------------------

The learning rule in equation \ref{eqn:learn_rule} is gated by a well-documented surprise signal \cite{Gerstner2018}. This surprise signal diffuses across the brain, and activates learning in the \textit{RNN}.

For each \textit{US} the following surprise signal is computed throughout the trial:
\begin{equation}
S^i(t) = \delta(t-t_\text{trig}) \, \bigl(\mathbbm{1}_{US^i} - E^i(t-t_\text{syn})\bigr),
\label{eqn:surprise}
\end{equation}
where $\mathbbm{1}_{US^i}$ is an indicator function for the presence of \textit{US-i}, $\delta$ is the Dirac delta function and $t_\text{trig}$ the time a surprise signal is triggered. In trials where the \textit{US} appears, we set $t_\text{trig}=t_\text{us-on}+t_\text{syn}$, where $t_\text{syn}=2*\tau_\text{s}=200 \, \textit{ms}$ is a synaptic transmission delay for the detection of the \textit{US} which matches well perceptual delays \cite{Picton1992}. The expectation $E^i$ also lags by the same amount, representing synaptic delays from the associative network to the surprise computation area. As can be seen in \cref{eqn:surprise}, the more the \textit{US} is expected upon its presentation, the lower the surprise. In extinction trials, we set $t_\text{trig}=t_\text{us-on}+t_\text{syn}+t_\text{wait}$, where $t_\text{wait}$ is a time after which a \textit{US} is no longer expected to arrive. The overall surprise signal is given by:
\begin{equation}
S = \sum_{i} S^i. 
\label{eqn:total_surprise}
\end{equation}

The surprise signal $S$ gives rise to neuromodulator release and uptake which determine the learning rate $\eta$. We assume that separate neuromodulators are at work for positive and negative surprise, and that they follow double-exponential dynamics \cite{Cragg2000}. 

Consider the case of positive surprise. The released and uptaken neuromodulator concentration $C^+_\text{r}$ and $C^+_\text{u}$ are given by:
\begin{equation}
\tau_\text{r} \frac {\textrm{d}C^+_\text{r}}{\textrm{d}t} = - C^+_\text{r} + \left[ S  \right]_+ 
\label{eqn:DA_released}
\end{equation}
and
\begin{equation}
\tau_\text{u} \frac {\textrm{d}C^+_\text{u}}{\textrm{d}t} = - C^+_\text{u} + C^+_\text{r}
\label{eqn:DA_uptaken}
\end{equation}
where $\tau_\text{r}$ and $\tau_\text{u}$ are the neuromodulator release and uptake time constants respectively, chosen to match the dopamine dynamics in \cref{fig:model}B in \cite{Cragg2000}. 

Negative surprise is controlled by a different neuromodulator, described by the following analogous dynamics:
\begin{equation}
\tau_\text{r} \frac {\textrm{d}C^-_\text{r}}{\textrm{d}t} = - C^-_\text{r} + \left[ - S  \right]_+
\end{equation}
and
\begin{equation}
\tau_\text{u} \frac {\textrm{d}C^-_\text{u}}{\textrm{d}t} = - C^-_\text{u} + C^-_\text{r}
\end{equation}

The neuromodulator uptake concentrations control the learning rate:
\begin{equation}
\eta = \eta_0 \ (C^+_\text{u} - C^-_\text{u}),
\end{equation}
where $\eta_0$ is the baseline learning rate.

% -----------------------------------------------------------
\subsection*{\textit{CS} short-term memory circuit}
% -----------------------------------------------------------

We now describe the short-term memory network used to maintain the $\hat{r}_\text{cs}$ representation that serves as input to the \textit{RNN}. 

To obtain a circuit that can maintain a short-term memory through persistent activity in the order of seconds \cite{Wang2001}, we train a separate recurrent neural network of point neurons using backpropagation through time (BPTT). These networks have been deemed to not be biologically plausible (although see \cite{Greedy2022}). However, for the purposes of our model we are only interested in the end product of a short-term memory circuit, and not in how the brain acquired such a circuit. Thus, BPTT provides an  efficient means of accomplishing this goal.

The memory circuit contains 64 neurons, and the vector of their firing rates $r_\text{mem}$ obeys:
\begin{equation}
\tau_\text{s} \frac {{\textrm d} r_\text{mem}}{{\textrm d}t} = - r_\text{mem} + \left [W_\text{mem} \, r_\text{mem} + W_\text{inp}  \, r_\text{cs} + b + n_\text{mem} \right ]_+
\label{eqn:RNN_dynamics}
\end{equation}
where $W_\text{mem}$ is a matrix with the connection weights between the memory neurons (dimension: $64 \times 64$), $W_\text{inp}$ is a matrix of connection weights for the incoming \textit{CS} inputs to the memory net (dimension: $64 \times N_\text{inp}$), $\tau_\text{s}$ is the same synaptic time constant described above, $b$ is a unit-specific bias vector, and $n_\text{mem}$ is a vector of IID Gaussian noise with zero mean and variance 0.01 added during training. A linear readout of the activity of the memory network provides the memory representation:
\begin{equation}
\hat{r}_\text{cs} = W_\text{out} \, r_\text{mem} ,
\label{eqn:RNN_output}
\end{equation}
where $W_\text{out}$ is a readout matrix (dimension: $N_\text{inp} \times 64$).

The weight matrices $W_\text{mem}$, $W_\text{inp}$, and $W_\text{out}$, as well as the bias vector $b$, are trained as follows. Every trial lasts for 3 seconds. On trial onset, a Boolean vector $r_\text{cs}$ is randomly generated and provided as input to the network. The \textit{CS} input is provided for a random duration drawn uniformly from $\left [ 0.5 , 2 \right ]$ seconds. The network is trained to output $r_\text{cs}$ at all times for trials that are 3 seconds long. We train the network for a total of $10^7$ trials in batches of 100. We use mean square error loss at the output, with a grace period 200 ms at the beginning of the trial where errors are not penalised. We optimise using Adam \cite{Kingma2014} with default parameters (decay rates for first and second moments $0.9$ and $0.999$ respectively, learning rate $0.001$). To facilitate BPTT, which does not scale well with the number of timepoints, we train the memory network using a time step of $10 * \Delta t$.

%% file: tables/table_1.tex
\begin{table}[!htb]
%\caption{\label{tab:params}Parameters values.}
% Use "S" column identifier to align on decimal point 
\begin{tabular}{c c c l}

\toprule
{Parameter} & Value          & Units                        & Description     \\
\midrule

$N_\text{stim}$ & $16$ &  & Number of \textit{CS-US}s pairs to be learnt  \\
$t_\text{trial}$ & $2$ & \textrm{s}  & Trial duration \\
$t_\text{cs-off}$ & $2$ & \textrm{s}  & Time in the trial at which \textit{CS} disappears  \\
$t_\text{us-on}$ & $1$ & \textrm{s}  & Time in the trial at which \textit{US} appears  \\

$N_\text{inp}$ & $20$ & & Stimuli input vector length \\
$H_\text{d}$ & $8$ &  & Minimal Hamming distance between behavioral stimulus vectors \\

$N_\text{rnn}$     & $64$ & & Number of associative neurons \\
$f_\text{max}$ & $100$ & \textrm{spikes/s} & Maximum firing rate\\
$\beta$ & $2$ & & Steepness of activation function\\
$V_{1/2}$ & $1.5$ & & Input level for 50 \% of the maximum firing rate \\
$\tau_\text{s}$     & $100$ &  \textrm{ms}         & Synaptic time constant     \\
$\tau_\text{l}$     & $20$ &  \textrm{ms}         & Leak time constant of dendritic compartment of associative neurons    \\
$C$     & $2$ &  \textrm{ms}         & Capacitance of somatic compartment of associative neurons    \\
$g_L$     & $0.1$ &  & Leak conductance of somatic compartment of associative neurons    \\
$g_D$     & $0.2$ &  &  Conductance from dendritic to somatic compartment    \\
$g_\text{inh}$ & $3/8$ & & Constant inhibitory conductance \\
$E_\text{e}$     & $14/3$ &  &  Excitatory synaptic reversal potential    \\
$E_\text{i}$     & $-1/3$ &  &  Inhibitory synaptic reversal potential    \\
$a$     & $0.95$ &  & Constant for deviation of the learning rule from self-consistency    \\

$\tau_\text{r}$ & $200$ & \textrm{ms} & Dopamine release time constant  \\
$\tau_\text{u}$ & $300$ & \textrm{ms} & Dopamine uptake time constant  \\
$\eta_0$ & $5*10^{-3}$ & & Baseline learning rate\\

$\Delta t$ & 1 & \textrm{ms} & Euler integration step size\\

\bottomrule
\end{tabular}

\medskip 

\textbf{Table 1: Model parameter values}. These values apply to all simulations, unless otherwise stated. Note that voltages, currents, and conductances are assumed unitless in the text; therefore capacitances have the same units as time constants.
\end{table}

%% file: supplementary.tex
%!TEX root = main.tex

\clearpage
% Reset numbering
%\runninglinenumbers[1]
\setcounter{page}{1}
\setcounter{table}{0}
\renewcommand{\thetable}{S\arabic{table}}
\setcounter{figure}{0}
\renewcommand{\thefigure}{S\arabic{figure}}

%\newcommand{\nextbest}{\max_{i \neq c} \mu_t^{(i)}}
%\newcommand{\postmean}{\mu_{t+1}^{(c)}}
%\newcommand{\currentmean}{\mu_{t}^{(c)}}
%\newcommand{\currentlam}{\lambda_{t}^{(c)}}
%\newcommand{\voiCDF}{\Phi \left( \frac{a - \mu_\mu}{\sigma_\mu} \right)}
%\newcommand{\conditionalx}{x_t \mid \currentmean, \currentlam}

% \newgeometry{left=1.2in, right=1.2in, top=1in, bottom=1in}
% \onecolumn

\section*{Supplemental Information}

% -----------------------------------------------------------
\subsection*{How does the \textit{RNN} learn?}
% -----------------------------------------------------------

In the main text, we've shown that the network is able to learn complex delay conditioning tasks using relatively few trials. In this section we explore in more detail the mechanisms through which the network solves the problem.

\Cref{fig:activations} shows how the activity of the associative neurons changes with training. It compares firing rates in response only to the \textit{CS}s (top), in response only to the associated \textit{US}s (middle), or in response to the full trial in which both are presented (bottom). Several things are worth noticing. 

First, the right column depicts the activity of the network after it has learnt the delay conditioning task. At this stage, the activity patterns in response to only the \textit{CS} or only the \textit{US} are very similar. This makes it possible to decode the upcoming \textit{US} using only the activity in the network in response to the associated \textit{CS}.

%%%%%%%%%%%
\input{figs_text/figS1_text.tex}
%%%%%%%%%%%

Second, the network learns mixed stimulus representations. This is important since there is evidence that the associative areas of the prefrontal cortex use this type of mixed coding \cite{Rigotti2013}.

Third, the pattern of activity in response to only the \textit{US} is unchanged by learning. This follows from the fact that in this case the response of the associative neurons is driven only by the input $r_\text{us}$ to the somatic compartment and the synaptic weights $W_\text{us}$ are not updated with training.

Fourth, the activity pattern in response to both the \textit{CS} and the \textit{US}, is very similar to the response to the \textit{US} alone, irrespective of the stage of learning. This is because the firing rate in our model is mainly controlled by the \textit{US}, while later in learning the \textit{CS} would induce the same response anyway. Overall, the learning rule modifies the \textit{CS} weights so that the \textit{CS} inputs are able to generate the representation of the \textit{US} both when the \textit{CS} is presented by itself, and when presented together with the \textit{US}.

\Cref{fig:trial_dynamics} provides further insight into the inner workings of the model. Each panel depicts the dynamics of a model component within a training trial. Columns denote different stages of training. Recall that the learning rule between associative neuron $i$ and input neuron $j$ is the product of three terms: a surprise modulated learning rate $\eta (S)$, the presynaptic potential in the input neuron $P_j$, and the neuron-specific firing rate error term $\left[\,f(V^\text{s}_i)-f(p^\prime \, V^\text{d}_i)\,\right]$. 

%%%%%%%%%%%
\input{figs_text/figS2_text.tex}
%%%%%%%%%%%

Consider the last term first. $f(V^\text{s}_i)$ is the firing rate of associative neuron $i$, which is determined by its somatic voltage $V^\text{s}_i$. $f(p^\prime \, V^\text{d}_i)$ is the (approximate) counterfactual firing rate that would occur if the \textit{CS} were presented by itself. When the \textit{US} is presented it dominates the activity of the associative neurons and thus the firing rate in the presence of both stimuli is similar to what would have been in the presence of only the \textit{US}. As a result, for the \textit{RNN} to be able to predict the \textit{US} in response to only the \textit{CS}, it has to be the case that $f(p^\prime \, V^\text{d}_i) \approx f(V^\text{s}_i)$. The learning rule implements a gradient like rule by increasing the \textit{CS} input weights when $f(p^\prime \, V^\text{d}_i) < f(V^\text{s}_i)$, and decreasing them when the opposite is true. As shown in the third row of \cref{fig:trial_dynamics}, these two variables are unrelated early in training, but converge to the same pattern as learning progresses.

Next consider the surprise modulated learning rate $\eta (S)$. Gating the learning rate by surprise is critical, as it provides a global reference signal crucial when there are more than one predictive \textit{CS}s available. Furthermore, biological neurons may not be able to compute $f(p^\prime \, V^\text{d}_i)$ exactly at the dendritic compartment, resulting in potential mismatches between $f(V^\text{s}_i)$ and $f(p^\prime \, V^\text{d}_i)$. If the learning rate $\eta$ were constant across training, these mismatches would result in slow unlearning when nothing behaviorally significant is happening. In contrast, when the learning rate is gated by surprise, the learning rate $\eta=0$ most of the times, and any mismatch between when $f(V^\text{s}_i)$ and $f(p^\prime  V^\text{d}_i)$ does not result in unlearning.

% OLD VERSION OF PREV PARAGRAPH // THINK
% Gating the learning rate by surprise is critical for extinction learning (see below). Once the model has learnt to predict the arrival of the \textit{US} in response to the \textit{CS}, that prediction stays active for several seconds, at least until there is sufficient decay of its short-term memory representation. As a result, in the absence of a surprise signal, the network has no way of learning that an expected \textit{US} has not arrived when expected. Since $S^i = - E^i$ in this case, the surprise modulated signal provides the necessary guidance.

Finally consider the presynaptic potential $P_j$. This term is present in most learning rules and reflects the old Hebbian dictum that "neurons that fire together wire together". In particular, other things being equal, the weights of more active synapse are updated more since they have a potentially stronger influence on the postsynaptic firing rate.

We emphasize again that the fact that associative neurons are two-compartment neurons is important for the biological plausibility of the model. The gradient like term $\left[\,f(V^\text{s}_i)-f(p^\prime \, V^\text{d}_i)\,\right]$ depends only on information available at the synapse, since it is based only on variables associated with that neuron. By definition, the presynaptic potential $P_j$ is also available at the synapse. Finally, the learning rate is implemented by neuromodulators that are diffused to the synapses of the associative network. As a result, all of the variables required to implement the learning rule are locally available at each synapse.

% -----------------------------------------------------------
\subsection*{Three-factor Hebbian learning fails at stimulus substitution}
% -----------------------------------------------------------

The results in the main text have shown that our model accounts for many common patterns in classical conditioning when the \textit{RNN} is trained with the predictive learning rule in equation \ref{eqn:learn_rule}. A key feature of this rule is that  learning is guided by a comparison of activity in the dendritic and somatic compartments of the associative neurons.

In this section we investigate the influence of the learning rule by asking whether the same network trained using previously proposed Hebbian plasticity rules is able to account for the same phenomena. To do this, we keep all of the model components unchanged except for the learning rule. We train the model with two widely used Hebbian-like plasticity rules, Oja's rule \cite{Oja1982} and the BCM rule \cite{Bienenstock1982}. We find that the resulting network either cannot learn multiple associations, or requires task specific parameter tuning.

Consider Oja's learning rule first. In this case the synaptic weights from input neuron $j$ to associative neuron $i$ are updated using the following rule:
\begin{equation}
\Delta W_{ij} = \eta (S) f(V^\text{s}_i) \left[\, P_j - n W_{ij} f(V^\text{s}_i) \right]\,  
\label{eqn:ojas_rule}
\end{equation}
where $n$ is the normalization strength. The normalization component is crucial, because otherwise learning would diverge. Normalization here focuses on the weights, and subjects the largest weights to the strongest normalization. We choose $n=40$ for which the final responses to the \textit{CS} span most of the output range of associative neurons in our model ($0-100 \, \textrm{spikes/s}$).

\Cref{fig:oja} shows the results of training the \textit{RNN} with this learning rule in the delay conditioning task. We find that the network can learn well when there is a single \textit{CS-US} pair, but it fails when it has to learn multiple associations. In fact, $r^\text{us-only}_\text{rnn}$ and $r^\text{cs-only}_\text{rnn}$ are anti-correlated in this case. This occurs because normalization introduces competition between incoming synapses to the same neuron \cite{Gerstner2014}, which in turn induces competition between the associations to be stored and leads to interference. More specifically, neurons that fire strongly for one pattern will sustain the harshest normalization in their incoming weights affecting the response to all other patterns. Fig. S4A also explores the role of the normalization coefficient and shows that its impact is minimal when learning a single association. This might be because final weight levels for active neurons are determined mostly by the firing rate $f(V^\text{s}_i)$ which is constant for the same association, and hence it serves as a modulator of the learning rate.

%%%%%%%%%%%
\input{figs_text/figS3_text.tex}
%%%%%%%%%%%

%Fig. S4 shows the results of training the \textit{RNN} with this learning rule in the delay conditioning task. We find that the network can learn well when there is a single \textit{CS-US} pair, and the result holds for a wide range of normalization values (Fig. S4a). This might be because final weight levels are determined mostly by the firing rate $f(V^\text{s}_i)$ which is constant for the same association, and hence it serves as a modulator of the learning rate. However, when having to learn multiple associations, Oja's rule fails. In this setting, normalization strength matters, and we pick $n=40$ which has the best performance. We find that $r^\text{us-only}_\text{rnn}$ and $r^\text{cs-only}_\text{rnn}$ are anti-correlated when learning multiple associations (Fig. S4b). This occurs because normalization introduces competition between incoming synapses to the same neuron \cite{Gerstner2014}, which in turn induces competition between the associations to be stored and leads to interference. More specifically, neurons that fire strongly for one pattern will sustain the harshest normalization in their incoming weights affecting the response to all other patterns. Fig. S4C explores the role of the normalization coefficient and shows that its impact is minimal when learning a single association.  

Now consider the BCM rule, which involves an alternative normalization strategy that, instead of focusing on the weights, sets a variable potentiation threshold for the postsynaptic firing rate. The rule is given by: 
\begin{equation}
\Delta W_{ij} = \eta (S) f(V^\text{s}_i) \left[\, f(V^\text{s}_i)  - \alpha  \theta_i \right]\, P_j  
\label{eqn:bcm_rule}
\end{equation}
where $\theta_i$ is a time-varying threshold, and $\alpha$  is a parameter that modulates the size of the threshold. A common choice is to make the threshold a function of the average recent firing rate, which we implement by making it an exponential moving average of the firing rate though the following differential equation:
\begin{equation}
\tau_\theta \frac {{\textrm d}\theta_i}{{\textrm d}t} = - \theta_i + f(V_i^\text{s})
\label{eqn:exp_avg}
\end{equation}
where the parameter $\tau_\theta$ determines the temporal window of integration. In theory, this approach sounds promising, since if $r^\text{cs-only}_{\text{rnn},i}<r^\text{both}_{\text{rnn},i}$, then $f(V_i^\text{s})>\theta_i$ leading to potentiation and vice versa, with this logic converging to $r^\text{cs-only}_{\text{rnn},i} \approx r^\text{both}_{\text{rnn},i}$. However, as we show this is not enough to guarantee the performance of the BCM rule.

\Cref{fig:bcm} shows the results of training the \textit{RNN} with the BCM rule and $\alpha = 1$. We find that with the same trial conditions used for our main results (as shown in \cref{fig:delay_cond}) the BCM rule generates intermediate amounts of conditioning, as it has a tendency to overshoot. Furthermore, \cref{fig:bcm}B shows that when we change the time at which the \textit{US} appears conditioning becomes even worse, and that the problem persists for different values of $\tau_{\theta}$. Since the BCM rule has a tendency of underestimating the impact of the \textit{CS}, we also explored a remedy that involved amplifying the threshold by setting $\alpha = 1.05 $. \Cref{fig:bcm}C shows that this can fix the problem for experiments in which $t_{us-on}= 1$ s, but as shown in \cref{fig:bcm}D, the performance of the network is still highly dependent on \textit{US} timing. This is because the threshold, determined by a moving averaging filter of the firing rate, is highly dependent on trial specifics. Therefore, we conclude that the time-dependent threshold of the BCM rule introduces sensitivity to experimental details that cannot be overcome.

%%%%%%%%%%%

\input{figs_text/figS4_text.tex}
%%%%%%%%%%%

Overall, the need to fine-tune the parameters of the BCM rule to specific trial details is a general problem of Hebbian learning rules, stemming from the fact that they lack supervision. A similar point has been made by \cite{Vafidis2022,Stringer2002}. In contrast, predictive learning does not demonstrate such sensitivity. \Cref{fig:bcm}E shows that the network learns the task well for a variety of \textit{US} onset times, without any explicit parameter tuning.

The results in this section showcase the importance of the predictive learning rule in this work, facilitated by the two-compartment nature of the associative neurons. The existence of two compartments, which separate \textit{CS} inputs to the dendritic compartment from \textit{US} inputs to the somatic compartment, makes it possible for the biologically plausible learning rule in eq. \ref{eqn:learn_rule} to compare the two and guide learning using only information locally available at the synapse. In this learning rule, the activity of the somatic compartment serves as a supervisory signal for learning the weights of the inputs to the dendritic compartment until they are able to fully predict their activity in response to the \textit{US}. In contrast, in this section we have shown that two canonical Hebbian rules struggle with this type of associative learning, in part because they do not have an analogous supervisory signal.

\clearpage

% -----------------------------------------------------------
\subsection*{Predictive coding and normative justification for the learning rule}
% -----------------------------------------------------------

In this section we provide further insight into the learning rule used in our model by showing that it follows directly from the objective of stimulus substitution. 

Stimulus substitution states that synaptic connections change during learning so that the activity of the associative network induced by the \textit{CS}  ($r^\text{cs-only}_\text{rnn}$) becomes identical to the response induced by the \textit{US} ($r^\text{us-only}_\text{rnn}$). It follows that the objective of stimulus substitution is to minimize the discrepancy or loss $\mathcal{L}$ between the two:
\begin{equation}
\mathcal{L} = \frac{1}{2} (r^\text{cs-only}_\text{rnn} - r^\text{us-only}_\text{rnn})^2
\label{eqn:stim_sub_error}
\end{equation}
We assume that the synaptic weights for \textit{US} inputs are fixed, since these are primary reinforcers. The synaptic weights for the \textit{CS} inputs are plastic, and they are shaped so that the \textit{CS} elicits the same response as the \textit{US}, essentially becoming predictive of the latter. Assuming a rectified linear (ReLU) activation function, $r^\text{cs-only}_\text{rnn}$ will obey
\begin{equation}
r^\text{cs-only}_\text{rnn} = \left[ \, W^\intercal P \, \right]_+
\label{eqn:CS_fr}
\end{equation}
where $W$ are the plastic synaptic weights for the \textit{CS} inputs, and $P$ are the postsynaptic potentialsof the input \textit{CS} neurons, low-pass filtered by synaptic delays. 

To minimize the loss $\mathcal{L}$, we perform local gradient descent with respect to $W$, which leads to the following update rule: 

\begin{equation}
\frac{\partial W}{\partial t} = - \eta \frac{\partial \mathcal{L}}{\partial W}.
\label{eqn:grad_desc}
\end{equation}
This results in the following update rule between input neuron $j$ and associative neuron $i$ from presynaptic neuron $j$:
\begin{equation}
\Delta W_{ij} = \eta \left(r^\text{us-only}_{\text{rnn},i} - r^\text{cs-only}_{\text{rnn},i} \right) P_j.
\label{eqn:norm_learn_rule}
\end{equation}

Here,  $r^\text{us-only}_{\text{rnn},i}$ acts as a "teacher" signal, in a setting that resembles self-supervised learning. Specifically, $r^\text{cs-only}_{\text{rnn},i}$ is compared to $r^\text{us-only}_{\text{rnn},i}$, and the discrepancy determines the sign and magnitude of weight change. However, only synapses from presynaptic neurons that have recently been active ($P_j>0$) are modified. This learning rule is said to perform predictive coding, because \textit{CS} inputs should predict (or anticipate) the response to the \textit{US}.

An implicit requirement of the learning rule is that there has to be a way to tell apart $r^\text{cs-only}_{\text{rnn},i}$ and $r^\text{us-only}_{\text{rnn},i}$, in order to compare them. However, a neuron only has a single output at a given time. Therefore, in principle it is unclear how the two firing rates could be compared in an online fashion and within the same neuron. The 2-compartment associative neurons resolve this because the activity in the somatic compartment $f(V^\text{s}_i)$ provides a measure of $r^\text{us-only}_{\text{rnn},i}$ \footnote{In reality, as we show in \cref{eqn:ss_volt} $V^\text{s}_i$ is affected by both somatic and dendritic inputs, however as we explain in the same section the influence of the dendritic inputs can never change the sign of $\left[\,f(V^\text{s}_i)-f(p^\prime \, V^\text{d}_i)\,\right]$, and the resulting weight changes are always in the correct direction.}, $f(p' V^\text{d}_i)$ provides a measure of $r^\text{cs-only}_{\text{rnn},i}$, and the information available to compute the former term is available in the dendritic compartment due to backpropagating action potentials \cite{Larkum1999}. Thus, the associative neurons contain all of the information needed to implement the learning rule that yields stimulus substitution.

\newpage

\section*{Supplementary Figures}

%%%%%%%%%%%%%%%%%%%%%%%%%%%%%%%%%%%%%%%%%

%%%%%%%%%%%
%\clearpage

\input{tables/table_S1.tex}

%%%%%%%%%%%

%%%%%%%%%%%
\clearpage
\input{figs_text/figS5_text.tex}
%%%%%%%%%%%

%%%%%%%%%%%
\clearpage
\input{figs_text/figS6_text.tex}
%%%%%%%%%%%

%%%%%%%%%%%
\clearpage
\input{figs_text/figS7_text.tex}
%%%%%%%%%%%

%% file: figs_text/figS1_text.tex
\begin{figure*}[t!]
  \centering
  \includegraphics[width=0.9\textwidth]{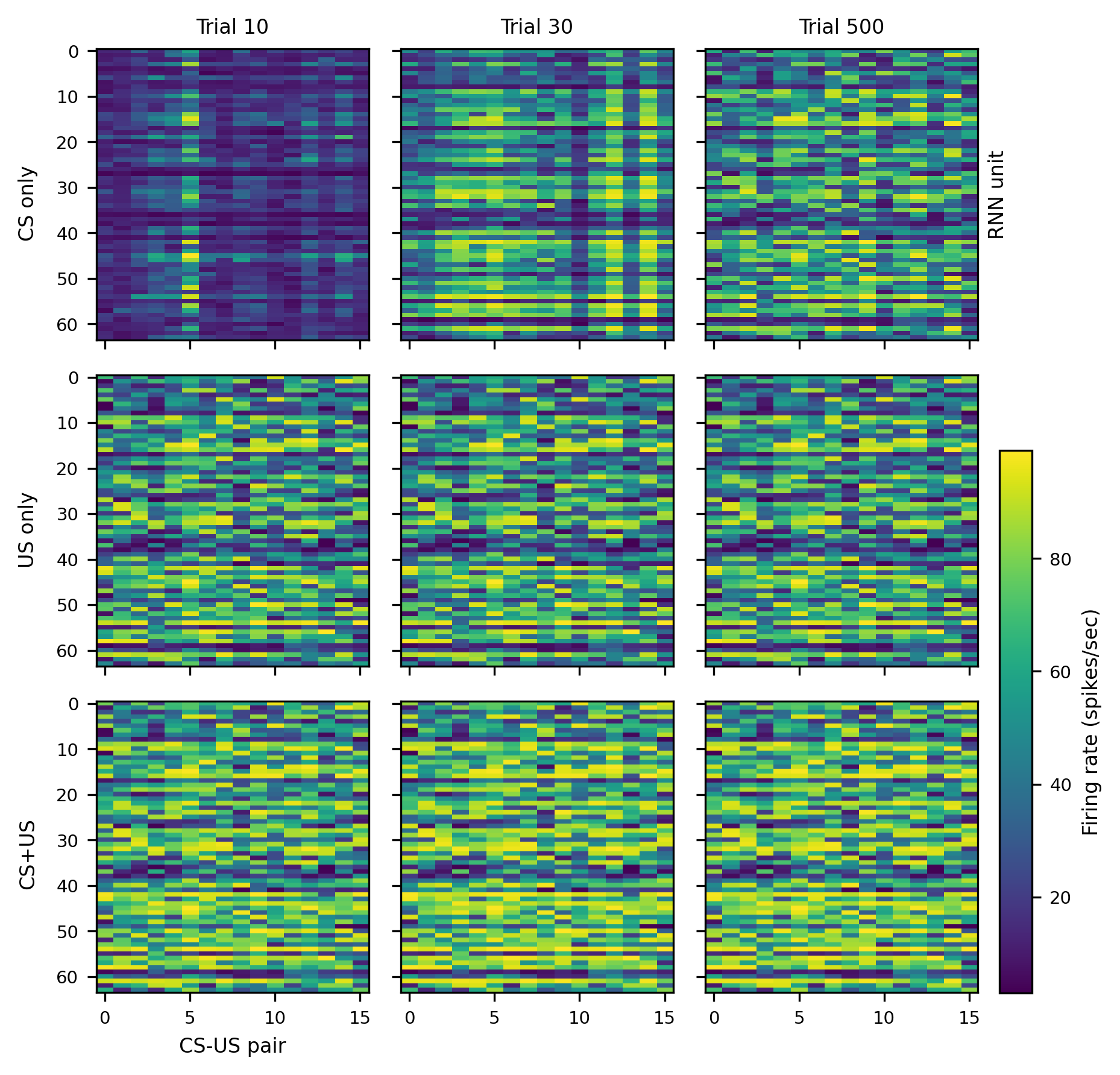}
  \caption{ \textbf{\textit{RNN} activity during learning}. 
  Firing rates in response to each stimulus pair at different stages of learning. Top row shows activity in response only to the associated \textit{CS}. Middle row shows activity in response only to the associated \textit{}{US}. Bottom row shows activity in response to the presentation of both. Activity is measured off-line (i.e., between learning trials).
 }
  \label{fig:activations}
\end{figure*}

%% file: figs_text/figS2_text.tex
\begin{figure*}[t!]
\centering
\includegraphics[width=0.9\textwidth]{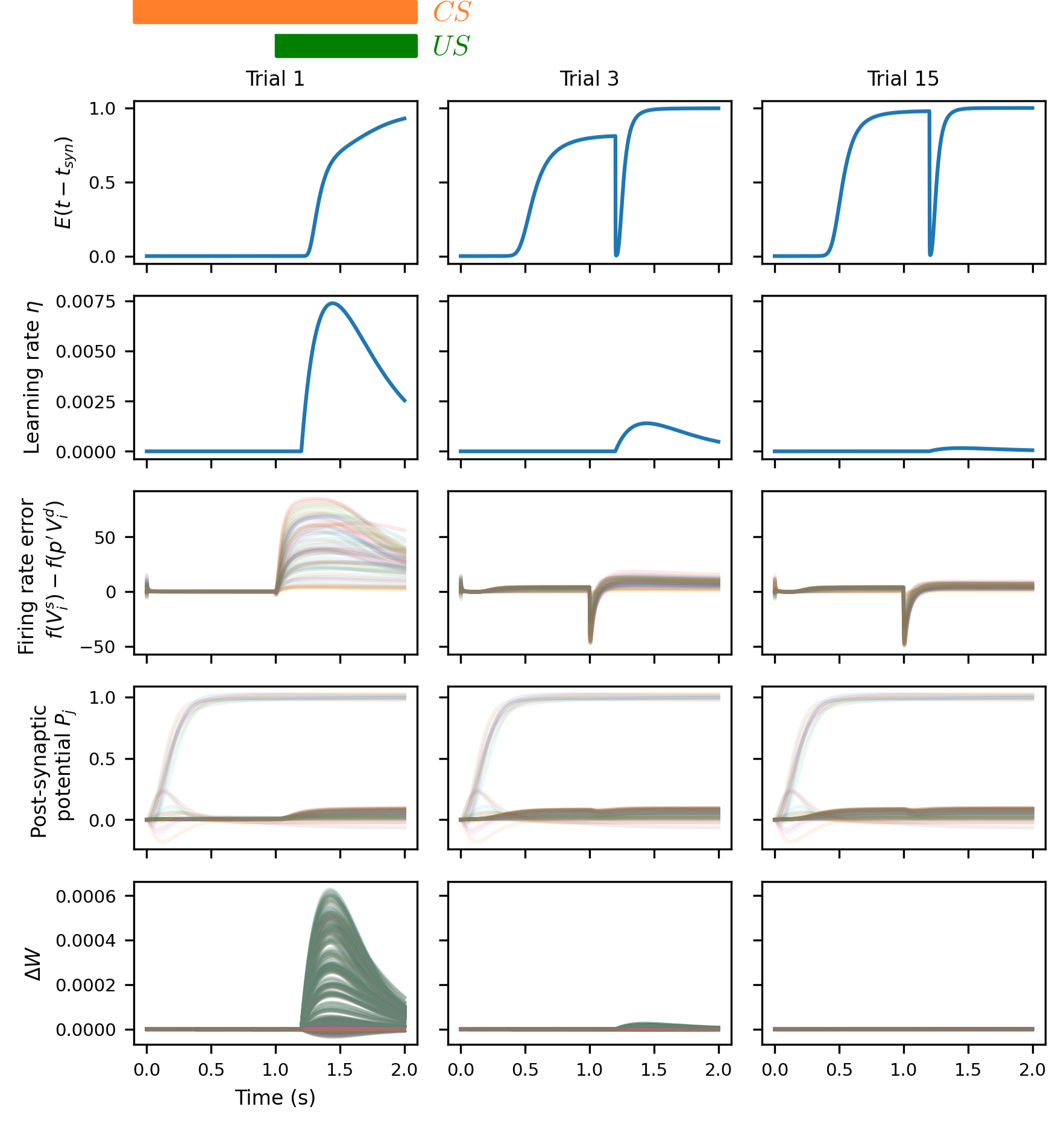}
\caption{\textbf{Within trial dynamics of model components}. Each panel depicts the dynamics of a model component within a training trial. Columns denote the level of training. Rows denote model variables. \( f(V^{\text{s}}_i) \) is the firing rate of associative neuron \( i \), which is determined by its somatic voltage \( V^{\text{s}}_i \). \( f(p' V^{\text{d}}_i) \) is the (approximate) counterfactual firing rate of the neuron when only the \textit{CS} is presented. \( E \) is the expectation signal for the \textit{US} shown in the trial. \( \eta(S) \) is the surprise-modulated learning rate. \( P_j \) is the presynaptic potentials of input neuron \( j \). \( \Delta W \) is the incremental weight change for elements of each element in \( W_{\text{rnn}} \) and \( W_{\text{cs}} \).}
\label{fig:trial_dynamics}
\end{figure*}

%% file: figs_text/figS3_text.tex
\begin{figure*}[t!]
  \centering
  \includegraphics[width=0.7\textwidth]{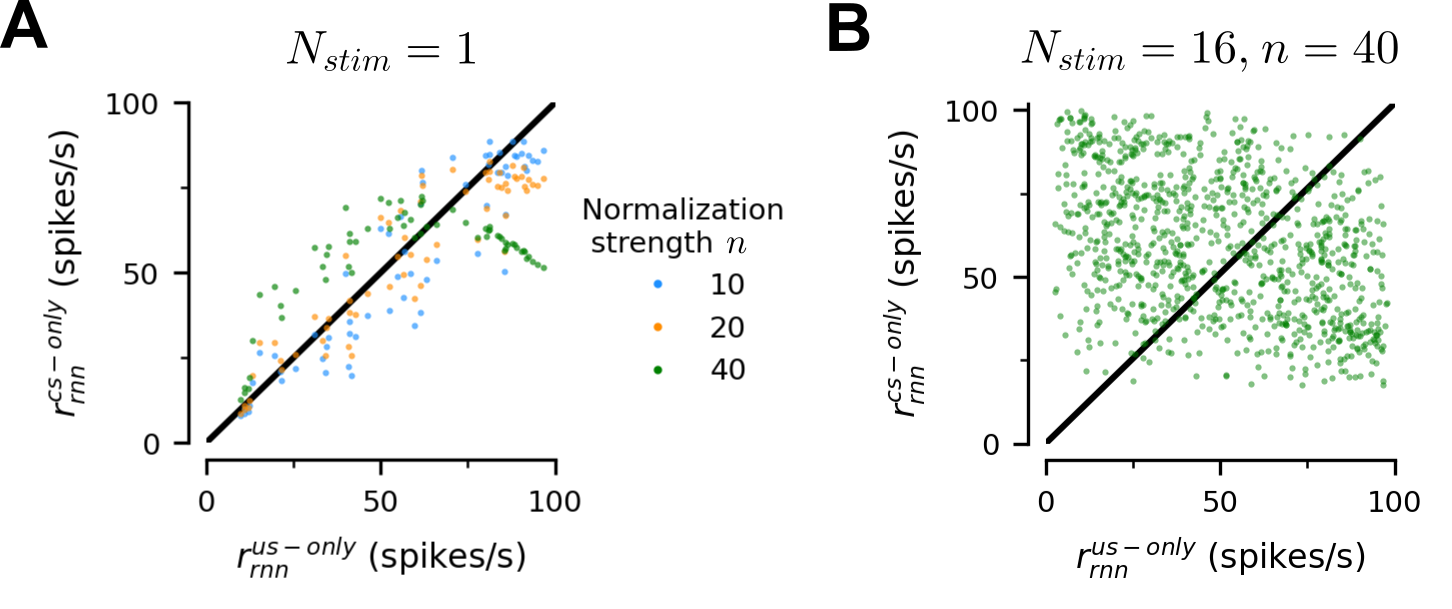}
  \caption{ \textbf{Delay conditioning with Oja's rule}. Each point compares the firing rate of an associative neuron at that stage of learning for a specific \textit{CS-US} pair when only the \textit{US}, or only the associated \textit{CS} are presented. Model is trained using Oja's rule. 
  (A) Model learns stimulus substitution for different normalization strengths when $N_\text{stim} = 1$. Model trained for 100 trials with $\eta_0=2*10^{-4}$.
  (B) Models fails to learn after 1000 training trials (64 per \textit{CS-US} pair) when $N_\text{stim} = 16$.  For this experiment, we use $\eta_0=10^{-3}$.
  }
  \label{fig:oja}
\end{figure*}

%% file: figs_text/figS4_text.tex
\begin{figure*}[t!]
  \centering
  \includegraphics[width=0.8\textwidth]{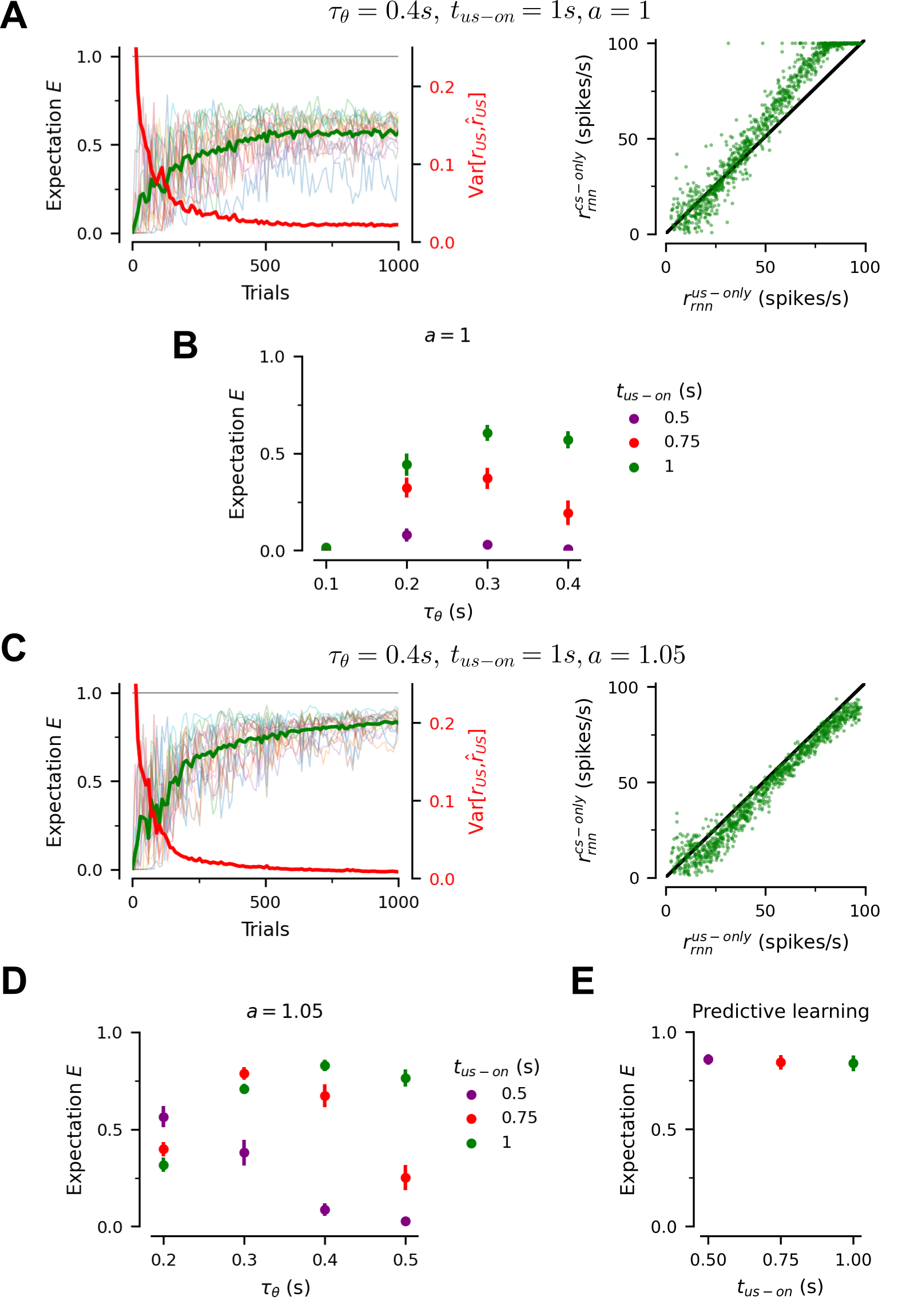}
  \caption{ \textbf{Delay conditioning with BCM rule}. Model trained with $\eta_0=0.3$ for 1000 trials in order to learn $N_\text{stim}=16$ associations. Model parameters and task conditions vary across panels.
  (A,C) Left: learning path. Green curve depicts the average expectation across \textit{CS-US}pairs. Learned expectations for individual pairs are shown in faint thin lines. Right: firing rates of all associative neurons after training.
  (B,D) Network performance, as measured by $E$, as a function of the parameter $\tau_{\theta}$ in the BCM rule and the timing at which the \textit{US} is presented.
  (E) Learnt \textit{US} expectations with our proposed predictive learning rule for different \textit{US} timings. In contrast to the BCM rule, predictive learning is insensitive to experimental details.
  }
  \label{fig:bcm}
\end{figure*}

%% file: tables/table_S1.tex
\begin{table}[!htb]
%\caption{\label{tab:params}Parameters values.}
% Use "S" column identifier to align on decimal point 
\begin{tabular}{c c c l}

\toprule
{Parameter} & Value          & Units                        & Justification     \\
\midrule

$N_\text{stim}$ & $16$ &  & Arbitrary, varied in text  \\
$t_\text{trial}$ & $2$ & \textrm{s}  & Arbitrary, varies across experiments \\
$t_\text{cs-off}$ & $2$ & \textrm{s}  & Arbitrary, varies across experiments  \\
$t_\text{us-on}$ & $1$ & \textrm{s}  & Arbitrary, varies across experiments  \\

$N_\text{inp}$ & $20$ & & Arbitrary \\
$H_\text{d}$ & $8$ &  & Arbitrary, varied in text \\

$N_\text{rnn}$     & $64$ & & Larger than number of associations learned \\
$f_\text{max}$ & $100$ & \textrm{spikes/s} & Matches plausible firing rates\\
$\beta$ & $2$ & & For smooth change of firing rates as activation varies\\
$V_{1/2}$ & $1.5$ & & Arbitrary \\
$\tau_\text{s}$     & $100$ &  \textrm{ms}         & Taken from literature     \\
$\tau_\text{l}$     & $20$ &  \textrm{ms}         & Taken from literature    \\
$C$     & $2$ &  \textrm{ms}         & Set by $\frac{\tau_\text{l}}{g_L}$    \\
$g_L$     & $0.1$ &  & Chosen to be similar in range as somatic conductances from inputs    \\
$g_D$     & $0.2$ &  &  Larger than $g_L$ to avoid large coupling attenuation and time lags    \\
$g_\text{inh}$ & $3/8$ & & Chosen to get unifrom firing rates distribution \\
$E_\text{e}$     & $14/3$ &  &  Taken from \cite{Urbanczik2014}    \\
$E_\text{i}$     & $-1/3$ &  &  Taken from \cite{Urbanczik2014}    \\
$a$     & $0.95$ &  & Arbitrary    \\

$\tau_\text{r}$ & $200$ & \textrm{ms} & Fit to experimental data \cite{Cragg2000}  \\
$\tau_\text{u}$ & $300$ & \textrm{ms} & Fit to experimental data \cite{Cragg2000}  \\
$\eta_0$ & $5*10^{-3}$ & & Chosen for fast learning when $N_\text{stim}=1$\\

$\Delta t$ & 1 & \textrm{ms} & Should be 10 times smaller than smallest time constant\\

\bottomrule
\end{tabular}

\medskip 

\textbf{Table S1: Parameter value justifications}. These values apply to all simulations, unless otherwise stated. Note that voltages, currents, and conductances are assumed unitless in the text; therefore capacitances have the same units as time constants.
\end{table}

%% file: figs_text/figS5_text.tex
\begin{figure*}[t!]
  \centering
  \includegraphics[width=0.8\textwidth]{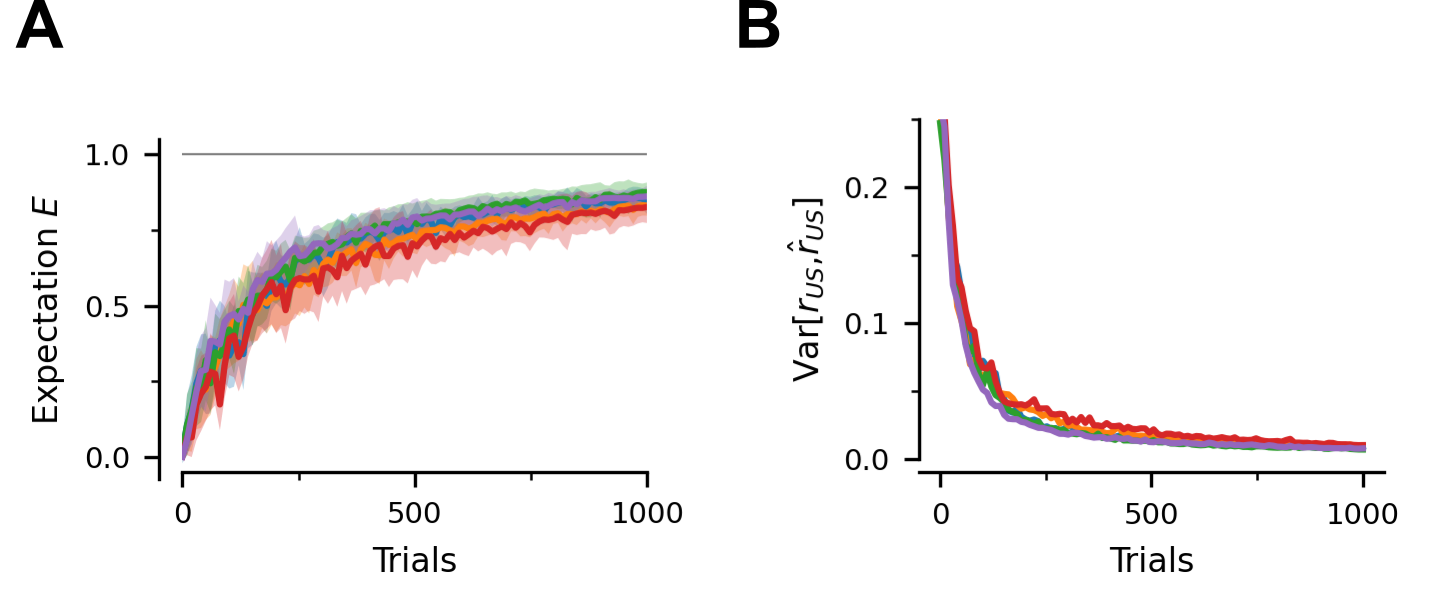}
  \caption{ \textbf{Variation across training runs}. Each curve depicts a different training run. Bands represent the $\mp$ SD across stimulus pairs.
  (A) Expectation for each \textit{US} after the network is presented only with the associated \textit{CS}, averaged across all pairs at different levels of training.
  (B) Distance between the true representation of the \textit{US}s ($r_\text{us}$) and their decoded representation $\hat{r}_\text{us}$ when presented only with the associated \textit{CS}, averaged across all pairs at different levels of training.
  }
  \label{fig:multiple_runs}
\end{figure*}

%% file: figs_text/figS6_text.tex
\begin{figure*}[t!]
  \centering
  \includegraphics[width=0.9\textwidth]{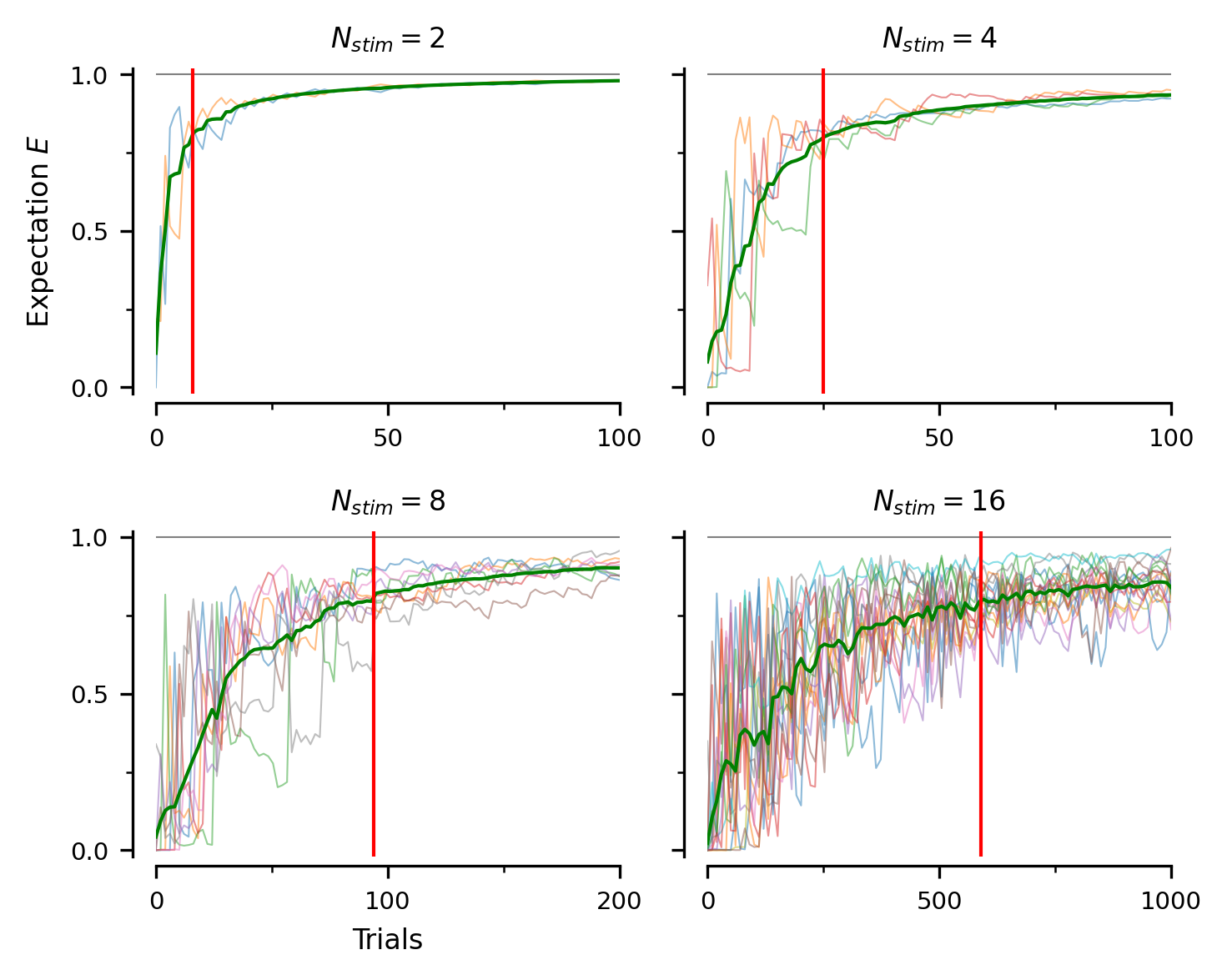}
  \caption{ \textbf{Impact of the number of stimulus pairs on delay conditioning}. Learning paths for each \textit{CS-US} pair for a single experimental run. Each thin line tracks the expectation $E$ for a single stimulus pair. Note that the paths do not increase monotonically, which shows that there can be interference across pairs. The vertical read lines indicate the time at which the average $E$ across pairs (thicker green line) reaches 80\% performance level.
  }
  \label{fig:n_stim}
\end{figure*}

%% file: figs_text/figS7_text.tex
\begin{figure*}[t!]
  \centering
  \includegraphics[width=0.9\textwidth]{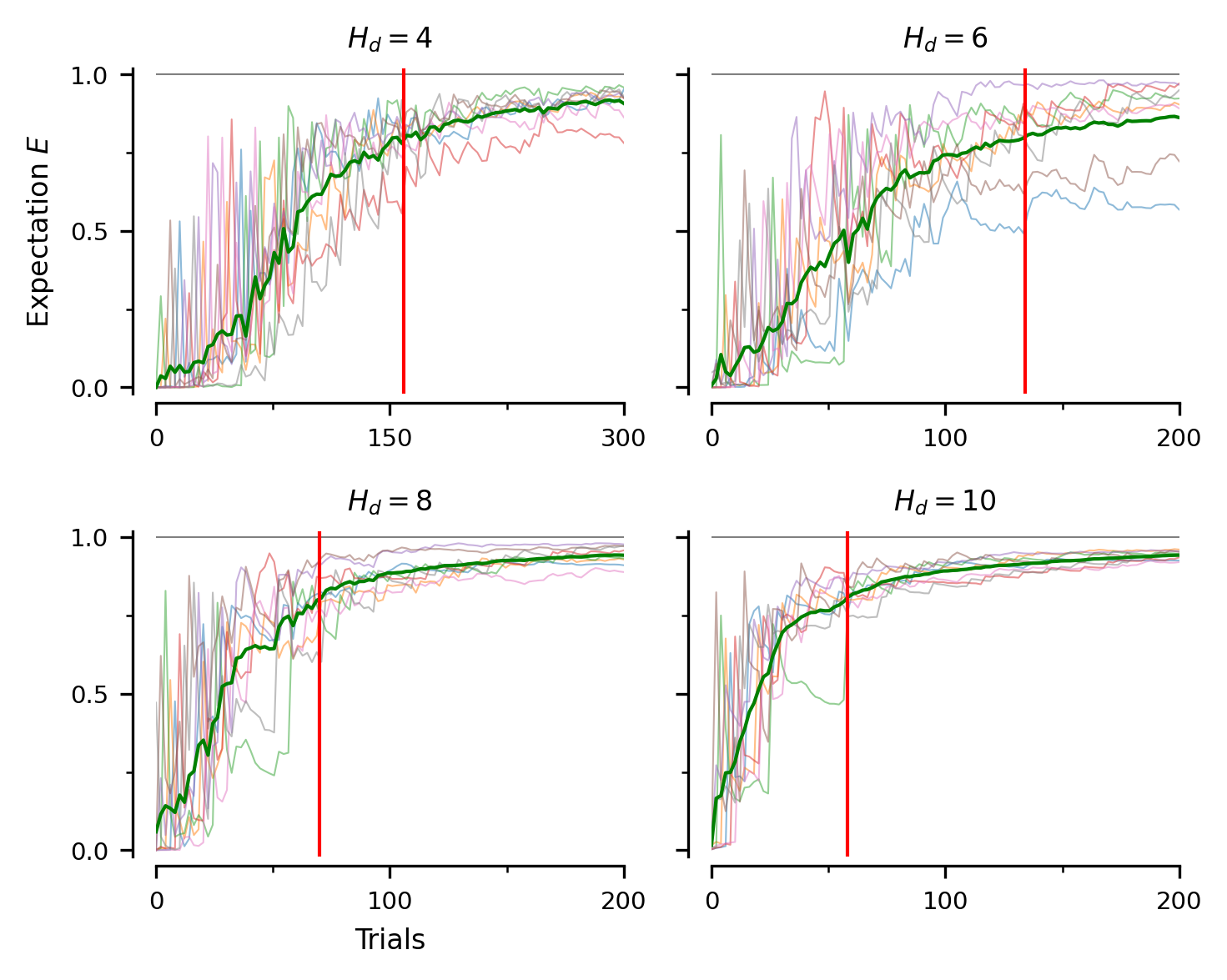}
  \caption{ \textbf{Impact of the similarity on stimulus representation on delay conditioning}. Learning paths for each \textit{CS-US} pair for a single experimental run. Each thin line tracks the expectation $E$ for a single stimulus pair. Note that the paths do not increase monotonically, which shows that there can be interference across pairs. The vertical read lines indicate the time at which the average $E$ across pairs (thicker green line) reaches 80\% performance level.
  }
  \label{fig:hamming}
\end{figure*}

%% file: main.bbl
\begin{thebibliography}{10}

\bibitem{Jenkins1973}
Jenkins HM, Moore BR.
\newblock The form of the auto-shaped response with food or water reinforcers.
\newblock Journal of the Experimental Analysis of Behavior. 1973;20(2):163--181.
\newblock doi:{10.1901/jeab.1973.20-163}.

\bibitem{Dai2023}
Dai J, Sun QQ. Learning induced neuronal identity switch in the superficial layers of the primary somatosensory cortex; 2023.
\newblock Available from: \url{http://dx.doi.org/10.1101/2023.08.30.555603}.

\bibitem{Sutton1987}
Sutton RS, Barto AG.
\newblock A temporal-difference model of classical conditioning.
\newblock In: Proceedings of the ninth annual conference of the cognitive science society. Seattle, WA; 1987. p. 355--378.

\bibitem{Klopf1988}
Klopf H.
\newblock A neuronal model of classical conditioning.
\newblock Psychobiology. 1988;16:85–--125.
\newblock doi:{10.3758/BF03333113}.

\bibitem{Balkenius1998}
Balkenius C, Morén J. Computational Models of Classical Conditioning: A Comparative Study; 1998.
\newblock Available from: \url{http://www.lucs.lu.se/People/Christian.Balkenius/PostScript/LUCS62.pdf}.

\bibitem{Izhikevich2007}
Izhikevich EM.
\newblock Solving the Distal Reward Problem through Linkage of {STDP} and Dopamine Signaling.
\newblock Cerebral Cortex. 2007;17(10):2443--2452.
\newblock doi:{10.1093/cercor/bhl152}.

\bibitem{Rigotti2013}
Rigotti M, Barak O, Warden MR, Wang XJ, Daw ND, Miller EK, et~al.
\newblock The importance of mixed selectivity in complex cognitive tasks.
\newblock Nature. 2013;497(7451):585--590.
\newblock doi:{10.1038/nature12160}.

\bibitem{Christian2003}
Christian KM, Thompson RF.
\newblock Neural Substrates of Eyeblink Conditioning: Acquisition and Retention.
\newblock Learning {\&} Memory. 2003;10(6):427--455.
\newblock doi:{10.1101/lm.59603}.

\bibitem{Gershman2021}
Gershman SJ, Balbi PE, Gallistel CR, Gunawardena J.
\newblock Reconsidering the evidence for learning in single cells.
\newblock {eLife}. 2021;10.
\newblock doi:{10.7554/elife.61907}.

\bibitem{Nieuwenhuys1994}
Nieuwenhuys R.
\newblock The neocortex. An overview of its evolutionary development, structural organization and synaptology.
\newblock Anatomy and Embryology. 1994;190(4).
\newblock doi:{10.1007/bf00187291}.

\bibitem{Doron2020}
Doron G, Shin JN, Takahashi N, Dr{\"u}ke M, Bocklisch C, Skenderi S, et~al.
\newblock Perirhinal input to neocortical layer 1 controls learning.
\newblock Science. 2020;370(6523):eaaz3136.
\newblock doi:{10.1126/science.aaz3136}.

\bibitem{Larkum2013}
Larkum M.
\newblock A cellular mechanism for cortical associations: an organizing principle for the cerebral cortex.
\newblock Trends in Neurosciences. 2013;36(3):141--151.
\newblock doi:{10.1016/j.tins.2012.11.006}.

\bibitem{Shin2021}
Shin JN, Doron G, Larkum ME.
\newblock Memories off the top of your head.
\newblock Science. 2021;374(6567):538--539.
\newblock doi:{10.1126/science.abk1859}.

\bibitem{Larkum1999}
Larkum ME, Zhu JJ, Sakmann B.
\newblock A new cellular mechanism for coupling inputs arriving at different cortical layers.
\newblock Nature. 1999;398(6725):338--341.
\newblock doi:{10.1038/18686}.

\bibitem{Urbanczik2014}
Urbanczik R, Senn W.
\newblock Learning by the Dendritic Prediction of Somatic Spiking.
\newblock Neuron. 2014;81(3):521--528.
\newblock doi:{10.1016/j.neuron.2013.11.030}.

\bibitem{Urbanczik2009}
Urbanczik R, Senn W.
\newblock Reinforcement learning in populations of spiking neurons.
\newblock Nature Neuroscience. 2009;12(3):250--252.
\newblock doi:{10.1038/nn.2264}.

\bibitem{Oja1982}
Oja E.
\newblock Simplified neuron model as a principal component analyzer.
\newblock Journal of Mathematical Biology. 1982;15(3):267--273.
\newblock doi:{10.1007/bf00275687}.

\bibitem{Bienenstock1982}
Bienenstock E, Cooper L, Munro P.
\newblock Theory for the development of neuron selectivity: orientation specificity and binocular interaction in visual cortex.
\newblock The Journal of Neuroscience. 1982;2(1):32--48.
\newblock doi:{10.1523/jneurosci.02-01-00032.1982}.

\bibitem{Gerstner2018}
Gerstner W, Lehmann M, Liakoni V, Corneil D, Brea J.
\newblock Eligibility Traces and Plasticity on Behavioral Time Scales: Experimental Support of {NeoHebbian} Three-Factor Learning Rules.
\newblock Frontiers in Neural Circuits. 2018;12.
\newblock doi:{10.3389/fncir.2018.00053}.

\bibitem{Schneiderman1964}
Schneiderman N, Gormezano I.
\newblock Conditioning of the nictitating membrane of the rabbit as a function of CS-US interval.
\newblock Journal of Comparative and Physiological Psychology. 1964;57(2):188–195.
\newblock doi:{10.1037/h0043419}.

\bibitem{Napier1992}
Napier RM, Macrae M, Kehoe EJ.
\newblock Rapid reacquisition in conditioning of the rabbit’s nictitating membrane response.
\newblock Journal of Experimental Psychology: Animal Behavior Processes. 1992;18(2):182–192.
\newblock doi:{10.1037/0097-7403.18.2.182}.

\bibitem{Gottlieb1998}
Gottlieb JP, Kusunoki M, Goldberg ME.
\newblock The representation of visual salience in monkey parietal cortex.
\newblock Nature. 1998;391(6666):481–484.
\newblock doi:{10.1038/35135}.

\bibitem{Rescorla1968}
Rescorla RA.
\newblock Probability of shock in the presence and absence of cs in fear conditioning.
\newblock Journal of Comparative and Physiological Psychology. 1968;66(1):1--5.
\newblock doi:{10.1037/h0025984}.

\bibitem{Hamblin1970}
Hamblin CL.
\newblock Fallacies.
\newblock London, England: Methuen young books; 1970.

\bibitem{Hopfield1982}
Hopfield JJ.
\newblock Neural networks and physical systems with emergent collective computational abilities.
\newblock PNAS. 1982;79:2554--2558.

\bibitem{Sompolinsky1986}
Sompolinsky H, Kanter I.
\newblock Temporal Association in Asymmetric Neural Networks.
\newblock Physical Review Letters. 1986;57(22):2861--2864.
\newblock doi:{10.1103/physrevlett.57.2861}.

\bibitem{Vasilaki2009}
Vasilaki E, Fr{\'{e}}maux N, Urbanczik R, Senn W, Gerstner W.
\newblock Spike-Based Reinforcement Learning in Continuous State and Action Space: When Policy Gradient Methods Fail.
\newblock {PLoS} Computational Biology. 2009;5(12):e1000586.
\newblock doi:{10.1371/journal.pcbi.1000586}.

\bibitem{Fremaux2010}
Fr{\'{e}}maux N, Sprekeler H, Gerstner W.
\newblock Functional Requirements for Reward-Modulated Spike-Timing-Dependent Plasticity.
\newblock Journal of Neuroscience. 2010;30(40):13326--13337.
\newblock doi:{10.1523/jneurosci.6249-09.2010}.

\bibitem{Brea2016}
Brea J, Ga{\'{a}}l AT, Urbanczik R, Senn W.
\newblock Prospective Coding by Spiking Neurons.
\newblock {PLoS} Computational Biology. 2016;12(6):e1005003.
\newblock doi:{10.1371/journal.pcbi.1005003}.

\bibitem{Wang2018}
Wang JX, Kurth-Nelson Z, Kumaran D, Tirumala D, Soyer H, Leibo JZ, et~al.
\newblock Prefrontal cortex as a meta-reinforcement learning system.
\newblock Nature Neuroscience. 2018;21(6):860--868.
\newblock doi:{10.1038/s41593-018-0147-8}.

\bibitem{Mante2013}
Mante V, Sussillo D, Shenoy KV, Newsome WT.
\newblock Context-dependent computation by recurrent dynamics in prefrontal cortex.
\newblock Nature. 2013;503(7474):78--84.
\newblock doi:{10.1038/nature12742}.

\bibitem{Thorndike1898}
Thorndike EL.
\newblock Animal intelligence: An experimental study of the associative processes in animals.
\newblock The Psychological Review: Monograph Supplements. 1898;2(4):i--109.
\newblock doi:{10.1037/h0092987}.

\bibitem{Rescorla1972}
Rescorla RA, Wagner AR.
\newblock A theory of Pavlovian conditioning: Variations in the effectiveness of reinforcement and nonreinforcement.
\newblock Current research and theory. 1972; p. 64--99.

\bibitem{Sutton1981}
Sutton RS, Barto AG.
\newblock Toward a modern theory of adaptive networks: Expectation and prediction.
\newblock Psychological Review. 1981;88(2):135--170.
\newblock doi:{10.1037/0033-295x.88.2.135}.

\bibitem{Lake2016}
Lake BM, Ullman TD, Tenenbaum JB, Gershman SJ.
\newblock Building machines that learn and think like people.
\newblock Behavioral and Brain Sciences. 2016;40:e253.
\newblock doi:{10.1017/s0140525x16001837}.

\bibitem{Fusi2016}
Fusi S, Miller EK, Rigotti M.
\newblock Why neurons mix: high dimensionality for higher cognition.
\newblock Current Opinion in Neurobiology. 2016;37:66–74.
\newblock doi:{10.1016/j.conb.2016.01.010}.

\bibitem{Vafidis2022}
Vafidis P, Owald D, D'Albis T, Kempter R.
\newblock Learning accurate path integration in ring attractor models of the head direction system.
\newblock eLife. 2022;11:e69841.
\newblock doi:{10.7554/eLife.69841}.

\bibitem{Fremaux2016}
Fr{\'{e}}maux N, Gerstner W.
\newblock Neuromodulated Spike-Timing-Dependent Plasticity, and Theory of Three-Factor Learning Rules.
\newblock Frontiers in Neural Circuits. 2016;9.
\newblock doi:{10.3389/fncir.2015.00085}.

\bibitem{Picton1992}
Picton TW.
\newblock The P300 Wave of the Human Event-Related Potential.
\newblock Journal of Clinical Neurophysiology. 1992;9(4):456–479.
\newblock doi:{10.1097/00004691-199210000-00002}.

\bibitem{Cragg2000}
Cragg SJ, Hille CJ, Greenfield SA.
\newblock Dopamine Release and Uptake Dynamics within Nonhuman Primate {StriatumIn} Vitro.
\newblock The Journal of Neuroscience. 2000;20(21):8209--8217.
\newblock doi:{10.1523/jneurosci.20-21-08209.2000}.

\bibitem{Wang2001}
Wang XJ.
\newblock Synaptic reverberation underlying mnemonic persistent activity.
\newblock Trends in Neurosciences. 2001;24(8):455--463.
\newblock doi:{10.1016/s0166-2236(00)01868-3}.

\bibitem{Greedy2022}
Greedy W, Zhu HW, Pemberton J, Mellor J, Ponte~Costa R.
\newblock Single-phase deep learning in cortico-cortical networks.
\newblock Advances in Neural Information Processing Systems. 2022;35:24213--24225.

\bibitem{Kingma2014}
Kingma D, Ba J.
\newblock Adam: A Method for Stochastic Optimization.
\newblock International Conference on Learning Representations. 2014;.

\bibitem{Gerstner2014}
Gerstner W, Kistler WM, Naud R, Paninski L.
\newblock Neuronal dynamics.
\newblock Cambridge, England: Cambridge University Press; 2014.

\bibitem{Stringer2002}
Stringer SM, Trappenberg TP, Rolls ET, de~Araujo IET.
\newblock Self-organizing continuous attractor networks and path integration: one-dimensional models of head direction cells.
\newblock Network: Computation In Neural Systems. 2002;13(2):217--242.

\end{thebibliography}
